\documentclass[aps,prb,twocolumn,showpacs,superscriptaddress,floatfix]{revtex4-1}
\usepackage{dcolumn}
\usepackage{amsmath}
\usepackage{graphicx,epsfig,psfrag}
\usepackage{amssymb}
\usepackage{natbib}
\usepackage{url}
\usepackage[breaklinks=true]{hyperref}
\hypersetup{
        colorlinks=true,
}
\usepackage{bookmark}
\renewcommand{\vec}[1]{{\bf #1}}

\begin{document}

\title{Numerical renormalization group calculation of impurity internal energy and 
specific heat of quantum impurity models}
 
\author{L. Merker}
\author{T. A. Costi}
\affiliation
{Peter Gr\"{u}nberg Institut and Institute for Advanced Simulation, 
Research Centre J\"ulich, 52425 J\"ulich, Germany}

\begin{abstract}
We introduce a method to obtain the specific heat 
of quantum impurity models via a direct calculation of the 
impurity internal energy requiring only the evaluation of local 
quantities within a single numerical renormalization group (NRG) 
calculation for the total
system. For the Anderson impurity model, we show that the 
impurity internal energy can be expressed as a
sum of purely local static correlation functions and a term that
involves also the impurity Green function. The 
temperature dependence of the latter 
can be neglected in many cases, thereby 
allowing the impurity specific heat, $C_{\rm imp}$, 
to be calculated accurately from local 
static correlation functions; specifically via 
$C_{\rm imp}=\frac{\partial E_{\rm ionic}}{\partial T} 
+ \frac{1}{2}\frac{\partial E_{\rm hyb}}{\partial T}$, where 
$E_{\rm ionic}$ and $E_{\rm hyb}$ are the energies of the (embedded) impurity and
the hybridization energy, respectively. The term involving the Green function
can also be evaluated in cases where its temperature dependence is non-negligible, adding an
extra term to $C_{\rm imp}$.
For the non-degenerate Anderson impurity model, we show by comparison 
with exact Bethe ansatz calculations that the results 
recover accurately both the 
Kondo induced peak in the specific heat at low temperatures as well 
as the high temperature peak due to the resonant level. 
The approach applies to multiorbital and multichannel Anderson 
impurity models with arbitrary local Coulomb interactions. An application
to the Ohmic two state system and the anisotropic Kondo model is also
given, with comparisons to Bethe ansatz calculations. The approach 
could also be of interest within other impurity solvers, for example, within 
quantum Monte Carlo techniques. 
\end{abstract}

\pacs{75.20.Hr, 71.27.+a, 72.15.Qm}
% new: diamagnetism and paramagnetism, 75.20.Hr
% new: electronic conduction in metals and alloys, 72.15.Qm
% keep: 71.27.+a Strongly correlated electron systems; heavy fermions

%Heat capacity of crystalline solids: 65.40.Ba
% entropy, thermodynamics, 05.70.-a
% removed: 73.20.-r Electron states at surfaces and interfaces
%removed: 71.30.+h 	Metal-insulator transitions and other electronic transitions 

\date{\today}

\maketitle

\section{Introduction} 
\label{sec:introduction}

Quantum impurity models play an important role in condensed matter physics,
for example, as models of transition metal and rare-earth impurities in metals \cite {Hewson1997} 
or two-level systems \cite{Zawadowski1980,Caldeira1983,Leggett1987,Weiss2008,Zarand2005} 
and qubits \cite{Loss1998} interacting with an environment 
or in describing the Kondo effect in nanoscale devices such as
molecular transistors, \cite{Park2002,Yu2004,Roch2009,Parks2010}
semiconductor quantum dots, \cite{Goldhaber1998,Cronenwett1998,Kretinin2011}
carbon nanotubes,\cite{Nygard2000} and magnetic ions such as Co \cite{Madhavan1998,Otte2008}
or Ce \cite{Li1998} adsorbed on surfaces. In addition, they appear as the effective
models within dynamical mean field theory (DMFT) treatments of strongly correlated 
electron systems, such as heavy fermions and transition metal oxides. 
\cite{Metzner1989,Georges1996,Kotliar2004,Vollhardt2012}
Hence, new approaches to calculate their dynamic, thermodynamic and transport 
properties are potentially of wide interest. 

The numerical renormalization group (NRG) method, 
\cite{Wilson1975,KWW1980a,KWW1980b,Bulla2008} in particular, 
has proven very successful for the study of quantum impurity models.
The method, described briefly in the next section, gives both the
thermodynamic, \cite{Wilson1975,KWW1980a,KWW1980b,Oliveira1981}
dynamic, \cite{Frota1986,Sakai1989,Costi1992,Bulla1998,Hofstetter2000,
Anders2005,Peters2006,Weichselbaum2007} and transport properties \cite{Costi1994} 
of quantum impurities. Thermodynamic properties, such as the specific heat, are of particular interest 
for bulk systems, such as dilute concentrations of transition metal or rare-earth ions in 
non-magnetic metals.\cite{Hewson1997} A measurement of the temperature 
dependence of the specific heat or susceptibility of such systems provides 
important information about their physical behavior, for example, whether such systems
exhibit Fermi liquid or non-Fermi liquid behavior at low temperature and thus
information about the nature of their low energy excitations.
\cite{Gonzalez-Buxton1998,Loehneysen2007}

The usual approach to calculating the specific heat of quantum impurity models within 
the NRG method consists of a two-stage procedure, \cite{KWW1980a,KWW1980b,Oliveira1981,Bulla2008} in which 
the Hamiltonians of the total system $H$ is first diagonalized, followed by a similar
diagonalization for the host Hamiltonian $H_{0}$. Here, $H=H_{\rm imp}+H_{\rm int}+H_{0}$ is the
Hamiltonian of a quantum impurity (described by $H_{\rm imp}$), interacting with a host 
(described by $H_{0}$) via the interaction term $H_{\rm int}$. From the 
eigenvalues of $H$ and $H_{0}$, the grand canonical partition functions $Z={\rm Tr}\; e^{-\beta H}$ and 
$Z_{0}= {\rm Tr}\; {e^{-\beta H_{0}}}$ and the corresponding thermodynamic potentials 
$\Omega(T) =- k_{\rm B}T\ln Z$ and $\Omega_{0}(T) =- k_{\rm B}T\ln Z_{0}$ are constructed, 
where $\beta=1/k_{\rm B}T$ is the inverse temperature. The impurity contribution to the 
specific heat, $C_{\rm imp}(T)$, is then obtained by subtraction via $C_{\rm imp}(T)=C(T)-C_{0}(T)$,
where $C(T)$ and $C_{0}(T)$ are the specific heats of the total system and of 
the host system, respectively,
\begin{eqnarray}
C(T) &=&-T\frac{\partial^{2}\Omega(T)}{\partial\;T^{2}}
=k_{\rm B}\beta^{2}\langle (H-\langle\; H\rangle)^{2} \rangle,\label{spec-ctotal}\\
C_{0}(T) &=&-T\frac{\partial^{2}\Omega_{0}(T)}{\partial\;T^{2}}
=k_{\rm B}\beta^{2}\langle (H_{0}-\langle\; H_{0}\rangle)^{2} \rangle\label{spec-chost}\\
C_{\rm imp}(T)&=&C(T)-C_{0}(T)\label{spec-imp}.
\end{eqnarray}

In this paper we present a new approach to the calculation of the
impurity internal energy and specific heat of quantum impurity models 
within the numerical renormalization group (NRG) method.
\cite{Wilson1975,KWW1980a,KWW1980b,Bulla2008} It relies on expressing
the impurity internal energy in terms of local quantities, and as such
is not restricted to the NRG but may be implemented within any
impurity solver that calculates such quantities. The main result of
this paper is the (approximate) expression for the impurity specific heat 
of the Anderson model (see Sec.~\ref{sec:new approach})
\begin{eqnarray}
C_{\rm imp}(T) &=& \frac{\partial E_{\rm ionic}}{\partial T}  
+ \frac{1}{2}\frac{\partial E_{\rm hyb}}{\partial T},\label{eq:spec-heat-main}
\end{eqnarray}
where $E_{\rm ionic}=\langle H_{\rm imp}\rangle$ and 
$E_{\rm hyb}=\langle H_{\rm int}\rangle$. The main advantages of this approach are
that, (i), Eq.~(\ref{eq:spec-heat-main}) involves only a first temperature 
derivative and is expected to be more accurate for numerical evaluations 
than Eqs.~(\ref{spec-ctotal})-(\ref{spec-imp}) which involve a second temperature derivative 
of the thermodynamic potential, or, the calculation of the total energy fluctuation,
(ii), the host contribution to the internal energy $\langle H_{\rm 0}\rangle$ 
has been analytically subtracted out (see Sec.~\ref{sec:new approach}), 
so only the diagonalization of $H$ is required,
(iii), only local static correlation functions appearing in $\langle H_{\rm imp}\rangle$ 
and $\langle H_{\rm int}\rangle$ are required, and, (iv), as we shall show, the
new approach is less sensitive to discretization effects of the host than the
usual approach which evaluates expectation values of extensive quantities.
We illustrate the method by applying it to the Anderson impurity model and 
we compare the results for specific heats with those from the conventional
NRG approach \cite{Oliveira1981,Costi1994,Costa1997} and with exact results 
from thermodynamic Bethe ansatz calculations. \cite{Tsvelick1982,Wiegmann1983,Okiji1983}

Early approaches to the specific heat of dilute Kondo systems used an equation of
motion decoupling scheme for the Kondo model \cite{Nagaoka1965} and expressed
the impurity internal energy in terms of the local ${\rm T}$-matrix. The results 
obtained for the specific heat within this approximation were inadequate, 
violating, for example, Fermi liquid properties at low temperatures. 
\cite{Bloomfield1967} A formally exact expression for the internal energy
of the Anderson model, in terms of the local self-energy and the local Green
function, was obtained by Kj\"{o}llerstr\"{o}m et. al.,
in Ref.~\onlinecite{Kjoellerstroem1966}. They evaluated the
specific heat in the low density limit (corresponding to a small occupation
of the local level) obtaining correct results obeying Fermi liquid theory in
this limit.
 
The most reliable approaches to specific heats of quantum impurity models are the
Bethe ansatz method for integrable models \cite{Tsvelick1982,Wiegmann1983,Okiji1983,Rajan1982,
Desgranges1985,Sacramento1991,Bolech2005} and the NRG method.  
An important aspect of the latter, allowing it to access thermodynamic properties
on all temperature scales down to $T=0$, is the use of a logarithmic grid to represent the
quasi-continuous spectrum $\omega \in [-D,+D]$ of the host system, $H_{0}$. Thus 
$\omega\rightarrow\omega_{n}=\pm D\Lambda^{-n}, n=0,1,\dots$, where the
parameter $\Lambda> 1$ achieves a separation of the many energy scales in $H_{0}$ and thus in
$H$ (see Sec.~\ref{sec:conventional}). A large $\Lambda\gg 1$ allows calculations to 
reach low temperatures in fewer steps within the iterative diagonalization procedure of 
the NRG, and, in addition, a large $\Lambda\gg 1$ reduces the size of the truncation 
errors at each step in this procedure. \cite{KWW1980a} However, for $\Lambda\gg 1$, 
specific heats (and also susceptibilities), 
calculated by using a standard logarithmic grid, exhibit discretization oscillations, 
especially at low temperatures.\cite{Oliveira1994} On the other hand, calculations
at smaller $\Lambda\lesssim 3$, with less severe discretization oscillations, 
are more prone to truncation errors. In order to be able to carry out accurate
calculations at all temperatures, using $\Lambda\gg 1$, an averaging over several
discretizations of the host degrees of freedom has been introduced which essentially
allows exact calculations to be carried out. \cite{Oliveira1994,Campo2005}
With this refinement, the NRG approach has been used 
extensively in calculations of specific heats of quantum impurity models, 
\cite{Costa1997} with applications to the two-impurity Kondo model 
\cite{Silva1996,Campo2004} and the two-channel Anderson models.\cite{Ferreira2012}

The paper is organized as follows. In Sec.~\ref{sec:conventional}, 
the Anderson impurity model is described, and the NRG is outlined together
with a brief description of how thermodynamic properties are conventionally
calculated within NRG (at $\Lambda\gg 1$). 
In Sec.~\ref{sec:new approach}, we describe our new approach to specific heats of
quantum impurity models,
%\changes
{
using the Anderson impurity model as an example 
(with some further details given in Appendix~\ref{sec:appendix band contribution}).
The availability of exact Bethe ansatz results for this model, 
\cite{Tsvelick1982,Wiegmann1983,Okiji1983} allows a detailed
evaluation of the accuracy of our new approach to specific heats.
} 
Results at zero and finite magnetic fields are presented in 
Sec.~\ref{sec:symmetric model results} for the symmetric Anderson model. These are
compared to both exact Bethe ansatz results and results obtained in the conventional NRG approach. 
Sec.~\ref{sec:asymmetric model results} contains results for the asymmetric model with comparisons
to corresponding Bethe ansatz calculations. The thermodynamic Bethe ansatz (TBA) equations for the
Anderson impurity model and the details of their numerical solution can be found 
in Appendix~\ref{sec:appendix tba calculations}. In Sec.~\ref{sec:other models} we present
the generalization to multichannel and multiorbital Anderson impurity models and
to dissipative two state systems. For the Ohmic case, results for specific heats are compared
to corresponding Bethe ansatz results for the equivalent anisotropic Kondo model (AKM). 
Section~\ref{sec:discussion and conclusions} summarizes the main results of this paper and 
discusses possible future applications.

\section{Model, method and conventional approach to thermodynamics}
\label{sec:conventional}

We consider the Anderson impurity model,\cite{Anderson1961}
described by the Hamiltonian  $$H  = H_{\rm imp} + H_{\rm 0} + H_{\rm int}.$$
The first term, 
$H_{\rm imp}  = \sum_{\sigma}\varepsilon_{d}d_{\sigma}^{\dagger}d_{\sigma} 
+ Un_{d\uparrow}n_{d\downarrow}$, describes the impurity with local level
energy $\varepsilon_{d}$ and onsite Coulomb repulsion $U$, the second term,
$H_{\rm 0} =  \sum_{k\sigma}\epsilon_{k}c_{k\sigma}^{\dagger}c_{k\sigma}$, 
is the kinetic energy of non-interacting conduction electrons with dispersion $\varepsilon_{k}$, 
and, the last term, $H_{\rm int}  =  \sum_{k\sigma}V_{k}(c_{k\sigma}^{\dagger}d_{\sigma}
+d^{\dagger}_{\sigma}c_{k\sigma})$, is the hybridization between the local level
and the conduction electron states, with $V_{k}$ being the hybridization matrix element.
We shall also consider the effect of a magnetic field of strength $B$ by adding a
term $H_{B}=-g\mu_{B}B\,S_{z}$ to $H$ where $S_{z}$ is the $z$-component of the total
spin (i.e., impurity plus conduction electron spin), $g$ is the electron $g$-factor, 
and $\mu_{\rm B}$ is the Bohr magneton. We choose units such that $g=\mu_{\rm B}=1$.

The NRG procedure consists of the following steps. First, the conduction electron
energies $-D\le \varepsilon_{k}\le D$, where $D$ is the half-bandwidth, are 
logarithmically discretized about the Fermi level $\varepsilon_{F}=0$, that is, 
$\epsilon_{k}\rightarrow \epsilon_{n} = \pm D\Lambda^{-n}, n=0,1,\dots$ 
where $\Lambda>1$ is a momentum rescaling factor. 
We shall also consider generalized discretizations defined by a parameter $z$, 
such that $\epsilon_{0}=\pm D$ and $\epsilon_{n} = \pm D\Lambda^{-n -(1-z)}, n=1,\dots$,
with $z=1$ recovering the usual discretization. 
For $\Lambda\gg 1$, discretization induced oscillations of period 
$\ln \Lambda$ can be 
eliminated by averaging results for several $z$ in $(0,1]$.\cite{Oliveira1994,Campo2005} 
Second, the operators $c_{n\sigma}, n=0,1,\dots$, are rotated to a new set 
$f_{n\sigma},n=0,1,\dots$, with $Vf_{0\sigma}=\sum_{n=0}^{\infty}V_{k_{n}}c_{n\sigma}$, 
such that the discretized conduction band  
$H_{\rm 0}=
\sum_{n=0\sigma}^{\infty}\pm E_{n}(z) c_{n\sigma}^{\dagger}c_{n\sigma}$, 
with, for example, $E_{n}(z)=\frac{1}{2}(1+\Lambda^{-1})D\Lambda^{-n}$ for $z=1$,
takes the tri-diagonal form $H_{\rm 0}\rightarrow \sum_{n=0\sigma}^{\infty}\tilde{\epsilon}_{n}(z)
f_{n\sigma}^{\dagger}f_{n\sigma} + \sum_{n=0\sigma}^{\infty}t_{n}(z)
(f_{n\sigma}^{\dagger}f_{n+1\sigma}+ 
f_{n+1\sigma}^{\dagger}f_{n\sigma})$ 
in the new basis. Finally, within this new basis, the sequence of truncated 
Hamiltonians $H_{m},\,m=0,1,\dots$, where
$H_{m} = H_{\rm imp} + H_{\rm hyb} + \sum_{n=0\sigma}^{m}\tilde{\epsilon}_{n}(z)
f_{n\sigma}^{\dagger}f_{n\sigma} 
+ \sum_{n=0\sigma}^{m-1}t_{n}(z)(f_{n\sigma}^{\dagger}f_{n+1\sigma}+
f_{n+1\sigma}^{\dagger}f_{n\sigma}),$
with $H_{\rm hyb}=V\sum_{\sigma}(f_{0\sigma}^{\dagger}d_{\sigma}+d_{\sigma}^{\dagger}f_{0\sigma})$,
is iteratively diagonalized by using the recursion relation 
$H_{m+1}=H_{m}+\sum_{\sigma}\tilde{\epsilon}_{m+1}(z)f_{m+1\sigma}^{\dagger}f_{m+1\sigma} +
\sum_{\sigma}t_{m}(z)(f_{m\sigma}^{\dagger}f_{m+1\sigma}+f_{m+1\sigma}^{\dagger}f_{m\sigma}).$
This procedure \cite{KWW1980a,KWW1980b,Bulla2008} yields the 
eigenstates $|p\rangle_{m}$ and 
eigenvalues $E_{p}^{m}$ on a decreasing set of energy scales 
$\omega_{m}(z)\sim t_{m}(z), m=0,1,\dots$. Since the number of states increases as $4^{m+2}$,
only the lowest states are retained for $m\ge m_{0}$, where typically $m_{0}\ge 4-5$.
%\changes
{
This is implemented either by, (i), specifying an approximately constant number
of states $N_{\rm keep}$ to retain at each $m\ge m_{0}$, and $m_{0}$ will be fixed by
the precise value of $N_{\rm keep}$, or, (ii), by specifying 
that only those states with rescaled energies 
$(E_{p}^{m}-E^{m}_{GS})/t_{m}(z) < e_{c}(\Lambda)$ be retained for $m\ge m_{0}$, for
some predefined $m_{0}$, where $E_{GS}^{m}$ is the (absolute) groundstate energy at iteration $m$ 
and $e_{c}(\Lambda)$ is $\Lambda$-dependent cut-off energy. 
Combining the information from all iterations then allows the calculation of thermodynamics on 
all temperature scales of interest.\cite{Costa1997,Oliveira1994}
For most of the results in this paper, we used the truncation scheme (ii) with $m_{0}=4-5$ and 
$e_{c}(\Lambda)=20\sqrt{\Lambda}$, similar to the choice in Ref.~\onlinecite{Costa1997}. Some
calculations using the truncation scheme (i) with $N_{\rm keep}=860$ were also carried out 
in Sec.~\ref{sec:dissipative two state}. Both schemes
were found to work well by comparison with exact Bethe ansatz calculations. Whereas in scheme (i), 
a fixed number, $N_{\rm keep}$, of levels is retained for all iterations $m\ge m_{0}$, in scheme (ii), the number of 
retained states, initially large for $m\lesssim m_{0}$ (typically several thousand), 
starts to decrease with increasing $m$, eventually saturating 
to a few hundred states at $m\gg m_{0}$ (e.g., for $\Lambda=4$). While in both schemes 
only the retained states of iteration $m$ are used to set up the  
Hamiltonian $H_{m+1}$ for the next iteration, all states of iteration $m$ 
are available, and are used, in practice, to calculate the thermodynamics.
}

The specific heat is calculated within the approach of Campo and Oliveira in 
Ref.~\onlinecite{Campo2005}, which we shall refer to as 
the ``conventional'' approach: For any temperature $T$, 
we choose the smallest $m$ such that
$k_{\rm B}T > t_{m}(z)$ and we use the eigenvalues of $H_{m}$ to evaluate the partition function
$Z_{m}(T)=\sum_{p}e^{-E_{p}^{m}/k_{\rm B}T}$. The 
expectation value $\langle H\rangle$ 
is then calculated, followed by $\langle(H-\langle H\rangle)^2\rangle$  and the 
specific heat $C(T)$  (in addition, the
thermodynamic potential $\Omega(T)=-k_{\rm B}T \ln Z_{m}(T)$ may also be calculated).
Calculations are carried out for several values of the $z$ parameter and then averaged.
In the calculations reported below, we choose $z=(2i-1)/2n_{z}, i=1,\dots,n_{z}$ with
$n_{z}=2,4$ or $8$. 
This procedure is repeated for the conduction band Hamiltonian $H_{\rm 0}$ to obtain the 
host contribution to the specific heat, $C_{0}(T)$. Finally, the impurity specific heat is obtained
via $C_{\rm imp}(T)=C(T)-C_{0}(T)$.
The above prescription works well for $\Lambda \ge 4$, since the use of large $\Lambda$ 
reduces the size of truncation errors during the iterative diagonalization of $H$ and 
$H_{\rm 0}$.\cite{KWW1980a} Furthermore, the use of large $\Lambda$, implies that the
highest states of $H_{m}$ have energies $\gg t_{m}(z)\sim T$ so that $Z_{m}(T)$ is 
a good approximation to the partition function of the infinite system at temperature $T$. 
In addition to the specific heat, we also calculate the impurity contribution to the entropy, 
$S_{\rm imp}(T)=S(T)-S_{0}(T)$, where $S(T)$ and $S_{0}(T)$ are the entropies for $H$ and $H_{\rm 0}$,
respectively, and 
\begin{eqnarray}
S(T) &=& -\frac{\partial\Omega}{\partial T} = k_{\rm B}\ln Z(T) + \langle H\rangle/T,\\
S_{\rm 0}(T) &=& -\frac{\partial\Omega_{\rm 0}}{\partial T} = k_{\rm B}\ln Z_{0}(T) + \langle H_{0}\rangle/T.
\end{eqnarray}

\begin{figure}
\includegraphics[width=\linewidth,clip]{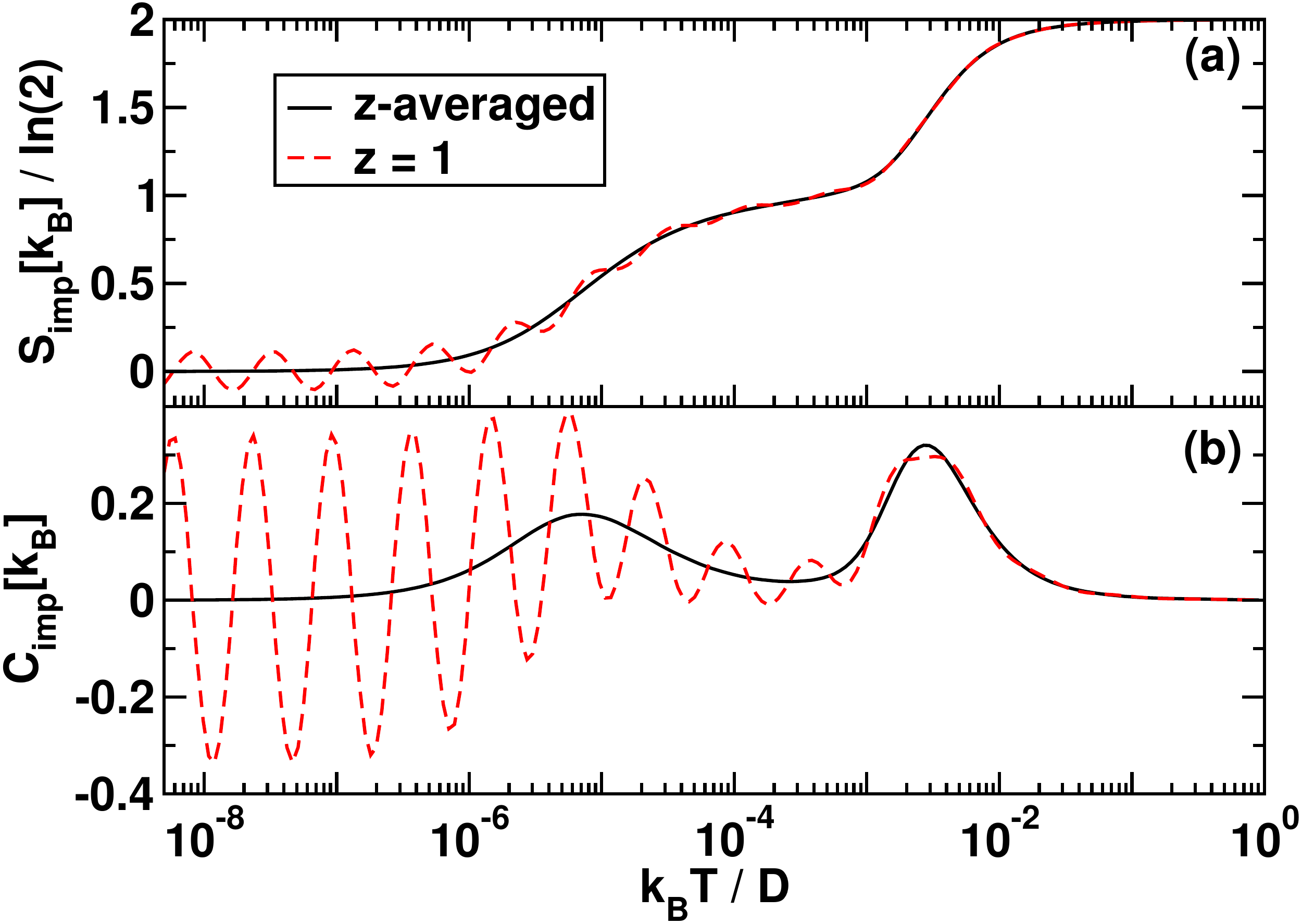}
\caption
{
  {\em (Color online)} 
  Temperature dependence of, (a), 
  the  impurity entropy, $S_{\rm imp}(T)$, 
  and, (b), 
  the impurity specific heat, $C_{\rm imp}(T)$,   
  for the symmetric Anderson
  model with $U/\Delta_{0}=12$ and $\Delta_{0}=0.001D$. The calculations are for $\Lambda = 4$ with an  energy 
  cut-off $e_{c}(\Lambda=4)=40$, without $z$-averaging [$n_{z}=1$, $z=1$ (dashed lines)], and
  with $z$-averaging [$n_{z}=2$, $z=1/4,\, 3/4$ (solid lines)]. For $\Lambda=4$ two $z$ values suffice to
  eliminate the discretization oscillations. 
  \label{fig1}
}
\end{figure}
Unless otherwise specified, the NRG calculations presented in this paper 
will be for a band of half-width $D=1$ and a constant particle-hole symmetric 
density of states $N_{F}=1/2D$. The hybridization strength, $\Delta_{0}$, defined as the half-width 
of the resonant level is given by $\Delta_{0}=\pi N_{F}V^{2}$. Calculations for the positive and 
negative-$U$ Anderson models include a ${\rm U}(1)$ symmetry for total electron number conservation 
and ${\rm SU}(2)$ symmetry for total spin conservation. We use the discretization scheme of 
Campo and Oliveira in Ref.~\onlinecite{Campo2005}.

\begin{figure}
\includegraphics[width=\linewidth,clip]{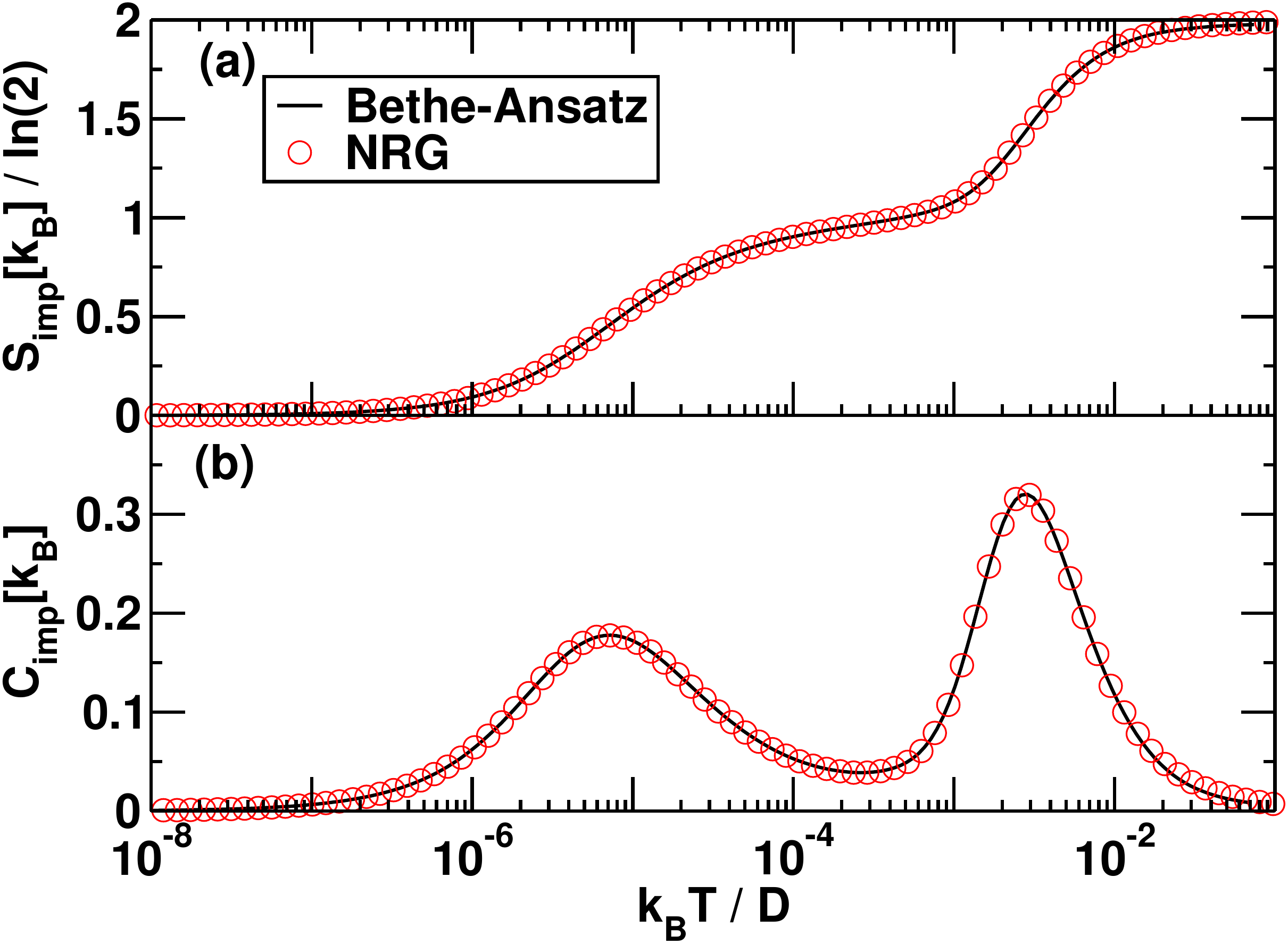}
\caption
{
  {\em (Color online)} 
  Temperature dependence of, (a), 
  the  impurity entropy, $S_{\rm imp}(T)$, 
  and, (b), 
  the impurity specific heat, $C_{\rm imp}(T)$, 
  for the symmetric Anderson
  model with $U/\Delta_{0}=12$ and $\Delta_{0}=0.001D$. 
  Symbols: NRG calculations using the conventional approach.
  Solid lines: Bethe ansatz calculations.
  The NRG calculations are $z$-averaged with $n_{z}=2$ and other parameters as in Fig.~\ref{fig1}.
  \label{fig2}
}
\end{figure}

Figure~\ref{fig1} shows the temperature dependence of the specific heat
and entropy, calculated with the above procedure, for the symmetric Anderson
model with $U/\Delta_{0}=12$ and $\Delta_{0}=0.001D$. 
The calculations are for 
$\Lambda = 4$ using an energy cut-off $e_{c}(\Lambda=4)=40$, both 
without $z$-averaging ($n_{z}=1$) and with $z$-averaging ($n_{z}=2$). Note the aforementioned
oscillations in the case of no $z$-averaging ($n_{z}=1$). For $\Lambda=4$,
two $z$ values suffice to eliminate the discretization oscillations 
(whereas for $\Lambda=10$, four values are required). In order to quantify the accuracy of
the NRG calculations, we also solved numerically the thermodynamic Bethe ansatz equations for the
Anderson model and calculated the entropy and specific heat (see 
Appendix~\ref{sec:appendix tba calculations} for details). A comparison of the $z$-averaged 
NRG calculations with the exact Bethe ansatz results, shown in Fig.~\ref{fig2}, indicates 
very good agreement. Nevertheless, in the next section we show that the specific heat 
can be calculated directly from the impurity contribution to the internal energy in terms 
of local static correlation functions and that discretization effects within this 
approach are less pronounced than those above.

\section{Impurity internal energy and specific heats}
\label{sec:new approach}

The impurity internal energy is defined by $E_{\rm imp}= E_{\rm total} - E_{0}$ where
$E_{\rm total}=\langle H\rangle$ and 
$E_{0} = \langle H_{0}\rangle=\sum_{k\sigma}\epsilon_{k}\langle 
c_{k\sigma}^{\dagger}c_{k\sigma}\rangle_{\rm 0}$, 
where the subscript $0$ denotes a thermodynamic
average for non-interacting conduction electrons (i.e., impurity is absent). We have
\begin{equation}
E_{0} = \sum_{\sigma}\int d\epsilon f(\epsilon)\epsilon N(\epsilon),
\end{equation}
where $f(\epsilon)$ is the Fermi function and
$N(\epsilon)=\sum_{k}\delta(\epsilon-\epsilon_{k})$ is the non-interacting
conduction electron density of states per spin. 
$E_{\rm total}$ has four contributions:
\begin{equation}
E_{\rm total} = E_{\rm occ} + E_{\rm docc} + E_{\rm cond} + E_{\rm hyb},
\end{equation}
where $E_{\rm occ} = \sum_{\sigma}\varepsilon_{d}\langle n_{d\sigma}\rangle$,
$E_{\rm docc} = U\langle n_{d\uparrow}n_{d\downarrow}\rangle$, 
$E_{\rm cond}=\sum_{k\sigma}\epsilon_{k}\langle c_{k\sigma}^{\dagger}c_{k\sigma}\rangle$ and
$E_{\rm hyb} = V\sum_{k\sigma}\langle c_{k\sigma}^{\dagger}d_{\sigma}
+d^{\dagger}_{\sigma}c_{k\sigma} \rangle$. The first two contributions are evaluated
as thermodynamic averages within the NRG calculation, requiring the calculation of matrix
elements of $\sum_{\sigma}n_{d\sigma}$ and the double occupancy operator 
$\hat{D}_{\rm occ}=n_{d\uparrow}n_{d\downarrow}$. The contribution $E_{\rm hyb}$ may also 
be evaluated as a thermodynamic average $E_{\rm hyb}= V\sum_{\sigma}\langle
d_{\sigma}^{\dagger}f_{0\sigma} + H.c.\rangle$. For the discussion below it
is useful to note that the contribution $E_{\rm hyb}$ can also be
expressed in terms of the local retarded d-electron Green function 
$G_{d\sigma}(\omega)=\langle\langle d_{\sigma}; d_{\sigma}^{\dagger}\rangle\rangle_{\omega+i\delta}$ 
and the hybridization function $\Delta(\omega) = \sum_{k}V^{2}/(\omega+i\delta-\epsilon_{k})$
as
\begin{equation}
E_{\rm hyb} = -\frac{2}{\pi}\sum_{\sigma}\int d\omega f(\omega) {\rm Im} \left[
G_{d\sigma}(\omega)\Delta(\omega)
\right].
\label{hyb-dynamics}
\end{equation}

\begin{figure}
\includegraphics[width=\linewidth,clip]{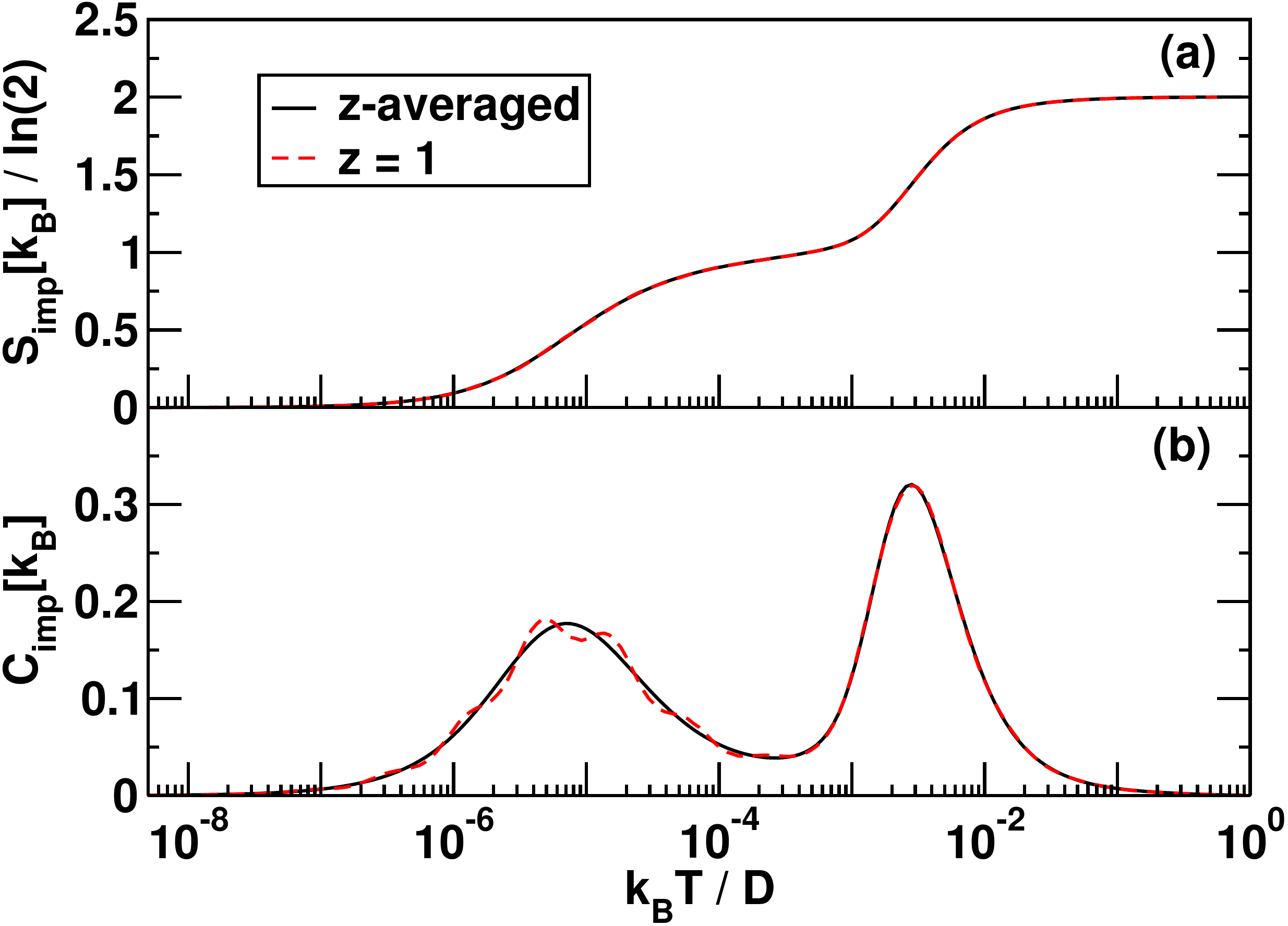}
\caption
{
  {\em (Color online)}
  Temperature dependence of, (a), the impurity entropy, $S_{\rm imp}(T)$, 
  and, (b), the impurity specific heat, $C_{\rm imp}(T)$, for the symmetric Anderson
  model with $U/\Delta_{0}=12$ and $\Delta_{0}=0.001D$ calculated within NRG using the new approach
  for $\Lambda = 4$ with an  energy cut-off $e_{c}(\Lambda=4)=40$. Solid lines: $n_{z}=2$ ($z$-averaging).
  Dashed lines: $n_{z}=1$ (no $z$-averaging). For $\Lambda=4$ two $z$ values thus suffice to
  eliminate the discretization oscillations at $n_{z}=1$.
  \label{fig3}
}
\end{figure}

Next, consider the contribution $E_{\rm cond}=
\sum_{k\sigma}\epsilon_{k}\langle c_{k\sigma}^{\dagger}c_{k\sigma}\rangle$. 
This is not simply $E_{0}$ since the impurity affects the conduction electrons once
$V$ is finite. It can be evaluated from the equation of motion of the retarded conduction electron
Greens function $G_{k\sigma}(\omega)=\langle\langle c_{k\sigma};c_{k\sigma}^{\dagger}\rangle\rangle_{\omega+i\delta}$:
\begin{equation}
G_{k\sigma} = G_{k\sigma}^{0}+ G_{k\sigma}^{0} {\cal T}_{\sigma}G_{k\sigma}^{0}.
\end{equation}
Here, ${\cal T}_{\sigma}(\omega)=V^{2}G_{d\sigma}(\omega)$ is the local ${\rm T}$-matrix and
$G_{k\sigma}^{0}(\omega)=1/(\omega+i\delta-\epsilon_{k})$ 
is the non-interacting conduction electron Greens function. Using
$$\langle c_{k\sigma}^{\dagger}c_{k\sigma} \rangle=-\frac{1}{\pi}\int d\omega f(\omega) {\rm Im} 
\left[\langle\langle
 c_{k\sigma};c_{k\sigma}^{\dagger}\rangle\rangle\right]$$
we find for $E_{\rm cond}$
$$
E_{\rm cond} = E_{0} + E_{\rm int},
$$
where
\begin{eqnarray}
E_{\rm int} &=& -\frac{1}{\pi}\sum_{\sigma}\int d\omega f(\omega)\int d\epsilon 
{\rm Im}\left[\frac{\epsilon V^{2}N(\epsilon)}
{(\omega+i\delta -\epsilon)^{2}}G_{d\sigma}(\omega)\right]\nonumber\\
 &=& -\frac{1}{\pi}\sum_{\sigma}\int d\omega f(\omega)
{\rm Im}\left[G_{d\sigma}(\omega)I(\omega)\right]\nonumber
\end{eqnarray}
where $I(\omega)$ is given by
$$I(\omega)=-\frac{1}{\pi}\int d\epsilon \frac{\epsilon \Delta_{\rm I}(\epsilon)}
{(\omega+i\delta -\epsilon)^{2}} = 
-\frac{\partial}{\partial\omega}\left(\omega\Delta(\omega)\right)
$$
with $\Delta_{\rm I}(\epsilon)={\rm Im}\left[\Delta(\epsilon+i\delta)\right]=-\pi V^{2}N(\epsilon)$,
and we evaluated $I(\omega)$ analytically by noting that 
$\Delta(\omega+i\delta)$ has the same properties as a retarded Green function
(see Appendix~\ref{sec:appendix band contribution} for details). We therefore find,
\begin{eqnarray}
E_{\rm int} &=& \frac{1}{\pi}\sum_{\sigma}\int d\omega f(\omega){\rm Im}
\left[G_{d\sigma}(\omega)
\frac{\partial}{\partial\omega}\left(\omega\Delta(\omega)\right)
\right]\nonumber\\
&=& E_{\rm int}^{(1)} + E_{\rm int}^{(2)},\label{band-contribution}\\
E_{\rm int}^{(1)} &=&\frac{1}{\pi}\sum_{\sigma}\int d\omega f(\omega){\rm Im}
\left[G_{d\sigma}(\omega)\Delta(\omega)\right]\\
E_{\rm int}^{(2)} &=&\frac{1}{\pi}\sum_{\sigma}\int d\omega f(\omega){\rm Im}
\left[G_{d\sigma}(\omega)\omega\frac{\partial\Delta(\omega)}{\partial\omega}\right]
\end{eqnarray}
From this and Eq.~(\ref{hyb-dynamics}) we see that $E_{\rm int}^{(1)}=-\frac{1}{2}E_{\rm hyb}$. 
Hence, the impurity contribution to the internal energy, 
$E_{\rm imp} = E_{\rm total}-E_{0}$, is given by
\begin{eqnarray}
E_{\rm imp} &=& E_{\rm occ} + E_{\rm docc} + \frac{1}{2}E_{\rm hyb} + E_{\rm int}^{(2)},\\
\label{eq:impurity-internal-energy-exact}
&=& E_{\rm ionic} + \frac{1}{2}E_{\rm hyb} + E_{\rm int}^{(2)},
\end{eqnarray}
where $E_{\rm ionic}=\langle H_{\rm imp}\rangle= E_{\rm occ} + E_{\rm docc}$ is 
adiabatically connected to the energy of the impurity decoupled from the band (i.e., its
energy at $V\rightarrow 0$).
All contributions to $E_{\rm imp}$, except for the last one, can be evaluated as thermodynamic
averages of local static correlation functions: The contribution $E_{\rm int}^{(1)}$ from 
the band which involves a finite frequency Greens function has been related to $E_{\rm hyb}$,
which can be evaluated as local static correlation function $V\sum_{\sigma}\langle d^{\dagger}_{\sigma}f_{0\sigma}+ H.c.\rangle$.
The contribution $E_{\rm int}^{(2)}$, also involves a finite frequency Greens function, 
%\changes
{
but we could not express this as a local static correlation function. 
}
Its temperature dependence, however, 
is negligible since the main temperature dependence arises from the Fermi window 
$|\omega|<T$, but this region is cut out in $E_{\rm int}^{(2)}$
due to the factor of $\omega$. In addition, for many cases of interest 
$\partial\left(\Delta(\omega)\right)/\partial\omega$ is small and vanishes in the
wide band limit: $D\rightarrow\infty$ and $\Delta_{0}=\pi N(0) V^{2}$ fixed. For example, 
for a constant density of states it equals $\frac{2\Delta_{0}}{\pi D}(1-(\omega/D)^{2})^{-1}\sim 
\Delta_{0}/D$ for $\omega\ll D$. Thus, to a very good approximation, which we shall quantify in the
rest of the paper with detailed numerical calculations and comparisons to exact Bethe 
ansatz results, we can approximate the impurity contribution to the specific heat and
entropy via $\bar{E}_{\rm imp} = E_{\rm ionic} + \frac{1}{2}E_{\rm hyb}$ as 
\begin{eqnarray}
\label{specific-heat-internal}
C_{\rm imp}(T) &=& \frac{\partial \bar{E}_{\rm imp}}{\partial T}
=\frac{\partial E_{\rm occ}}{\partial T} + \frac{\partial E_{\rm docc}}{\partial T} 
+ \frac{1}{2}\frac{\partial E_{\rm hyb}}{\partial T}\nonumber\\
&=&\frac{\partial E_{\rm ionic}}{\partial T}  
+ \frac{1}{2}\frac{\partial E_{\rm hyb}}{\partial T},\label{eq:impurity-spec-from-energy}\\
\label{entropy-internal}
S_{\rm imp}(T) &=& \int_{0}^{T}dT' \frac{C_{\rm imp}(T')}{T'}.
\end{eqnarray}
The omitted term, $\partial E_{\rm int}^{(2)}/\partial T$, in (\ref{eq:impurity-spec-from-energy}) 
as argued above, has a negligible  temperature dependence (although its magnitude is 
not necessarily always small compared to the terms retained). 
Notice that $\bar{E}_{\rm imp}$ is made up of a 
term due to the partial occupation of the local resonant level ($E_{\rm occ}$), 
a term due to the Coulomb repulsion of electrons in this level ($E_{\rm docc}$), and,
a term due to the energy gained by hybridization of the local level with the conduction
electrons ($E_{\rm hyb}/2$), that is, it involves only local static correlation functions.
Such quantities can be calculated very accurately and efficiently within the 
NRG method, within a single calculation for the total system only, a significant 
advantage of this approach. 
In some situations, the hybridization function $\Delta(\omega)$ may be strongly asymmetric and
have a strong energy dependence close to $\omega=0$. In such cases, the term $E_{\rm int}^{(2)}$
can be calculated via the local spectral function and included in $E_{\rm imp}$, which
is possible within the NRG, at somewhat higher numerical cost. 
Another advantage of the present approach, is that discretization oscillations are 
far smaller for local quantities appearing in $\bar{E}_{\rm imp}$ than for extensive 
quantities, such as $\langle H\rangle$ and $\langle (H-\langle H\rangle)^2\rangle$ 
appearing in the conventional approach to specific heats. 
Figure~\ref{fig3} shows the specific heat and entropy calculated with the above
%\changes
{
method, for the same parameters as in Fig.~\ref{fig1}-\ref{fig2}, with and without $z$-averaging. 
One sees that the discretization oscillations in the case of no $z$-averaging ($n_{z}=1$ curves) are 
drastically smaller than for the corresponding $n_{z}=1$ results from the conventional approach in Fig.~\ref{fig1}. 
Including $z$-averaging makes the results of the new procedure indistinguishable from the Bethe ansatz calculations,
as will be discussed in detail in  Sec.~\ref{sec:symmetric model results}-\ref{sec:asymmetric model results}.
}

\begin{figure}
\includegraphics[width=\linewidth,clip]{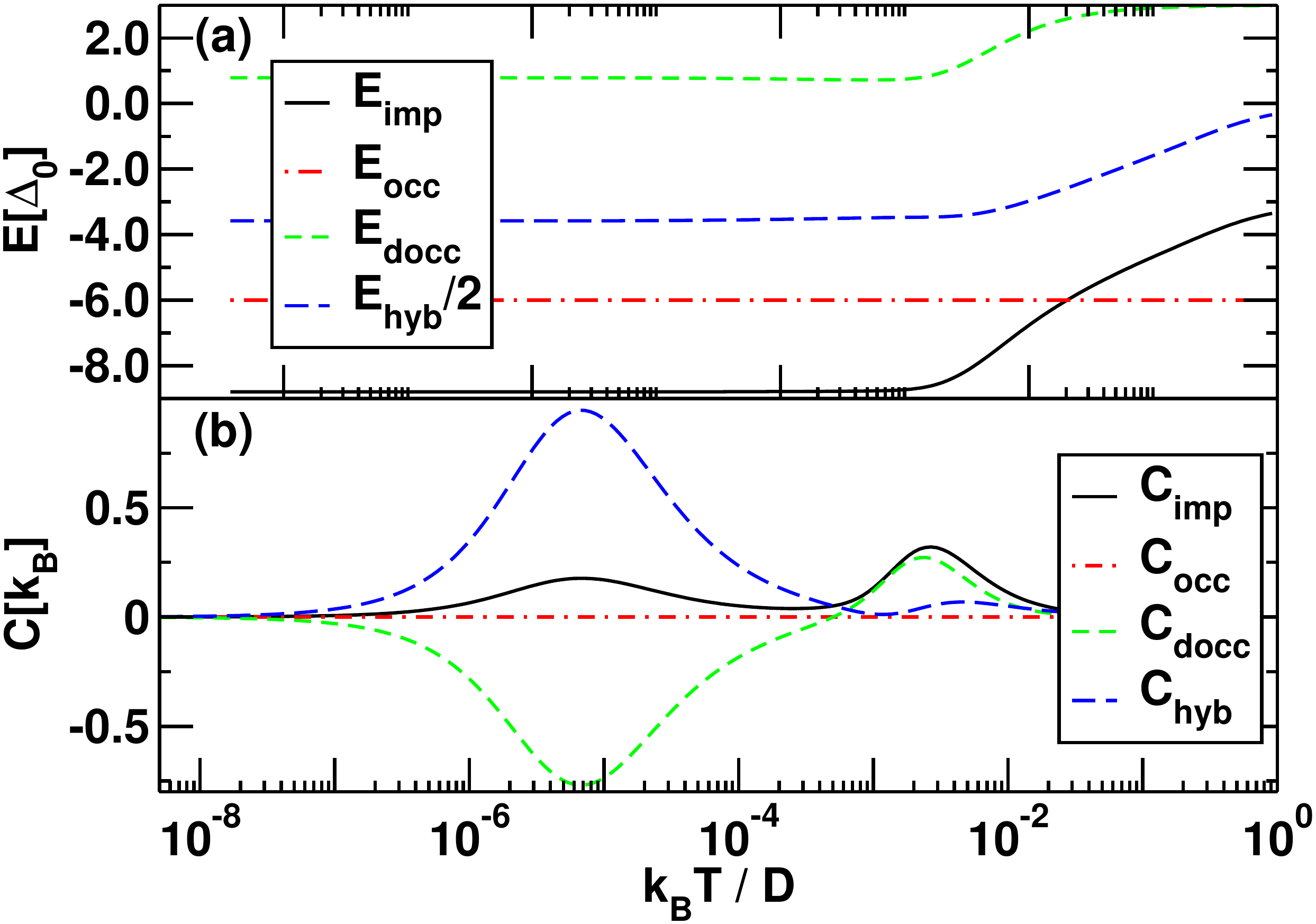}
\caption
{
  {\em (Color online)} 
  (a) The individual contributions
  $E_{\rm occ}$, $E_{\rm docc}$, and $E_{\rm hyb}$ to $E_{\rm imp}$ as a function of temperature
  (in units of $\Delta_{0}$) for the symmetric model with parameters as in Fig.~\ref{fig1}
  ($z$-averaged with $n_{z}=2$).
  (b) Temperature derivatives of the above, yielding the relative contributions 
  $C_{\rm occ}$, $C_{\rm docc}$, and $C_{\rm hyb}$ to the  specific heat $C_{\rm imp}$.
  \label{fig4}
}
\end{figure}

\begin{figure}
\includegraphics[width=\linewidth,clip]{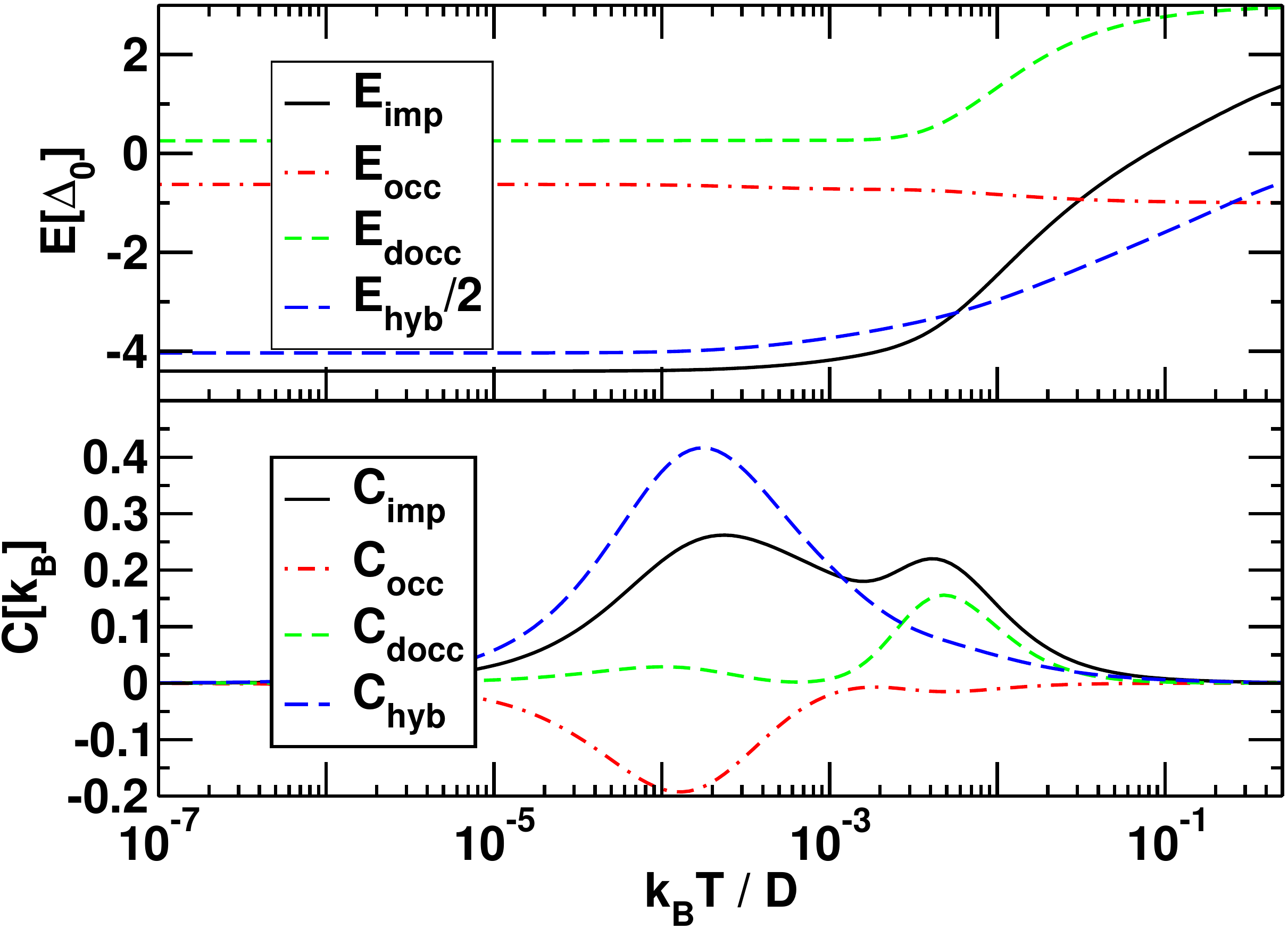}
\caption
{
  {\em (Color online)} 
  (a) The individual contributions
  $E_{\rm occ}$, $E_{\rm docc}$, and $E_{\rm hyb}$ to $E_{\rm imp}$ as a function of temperature
  (in units of $\Delta_{0}$) for the asymmetric model with parameters as in Fig.~\ref{fig1}, but
  for an asymmetric level position $\varepsilon_{d}/\Delta_{0}=-1$ ($z$-averaged with $n_{z}=2$).
  (b) Temperature derivatives of the above, yielding the relative contributions 
  $C_{\rm occ}$, $C_{\rm docc}$, and $C_{\rm hyb}$ to the  specific heat $C_{\rm imp}$.
  \label{fig5}
}
\end{figure}

In Fig.~\ref{fig4}(a), we show the different contributions $E_{\rm occ}$, 
$E_{\rm docc}$ and $E_{\rm hyb}/2$ to the impurity internal energy for the symmetric
Anderson model. Their temperature derivatives $C_{\rm occ}$ , $C_{\rm docc}$  and $C_{\rm hyb}$ 
give the relative contributions of these terms to the impurity specific heat $C_{\rm imp}$ 
and are shown in Fig.~\ref{fig4}(b). Notice, that the Kondo induced peak in $C_{\rm imp}$ 
at low temperatures results from a delicate balance of the hybridization ($C_{\rm hyb}$) 
and Coulomb contributions  ($C_{\rm docc}$), while the peak due to the resonant level at
high temperatures is mainly due to the Coulomb term. The latter trend persists also for the 
asymmetric model, as shown in Fig.~\ref{fig5}. Notice also that the gain
in energy due to hybridization diminishes at high temperatures, reflecting the 
decoupling of the impurity from the conduction electrons in this limit. In general,
however, the interaction of the impurity with the environment via the hybridization term
provides an essential contribution at all non-zero hybridization strengths. 

\begin{figure}
\includegraphics[width=\linewidth,clip]{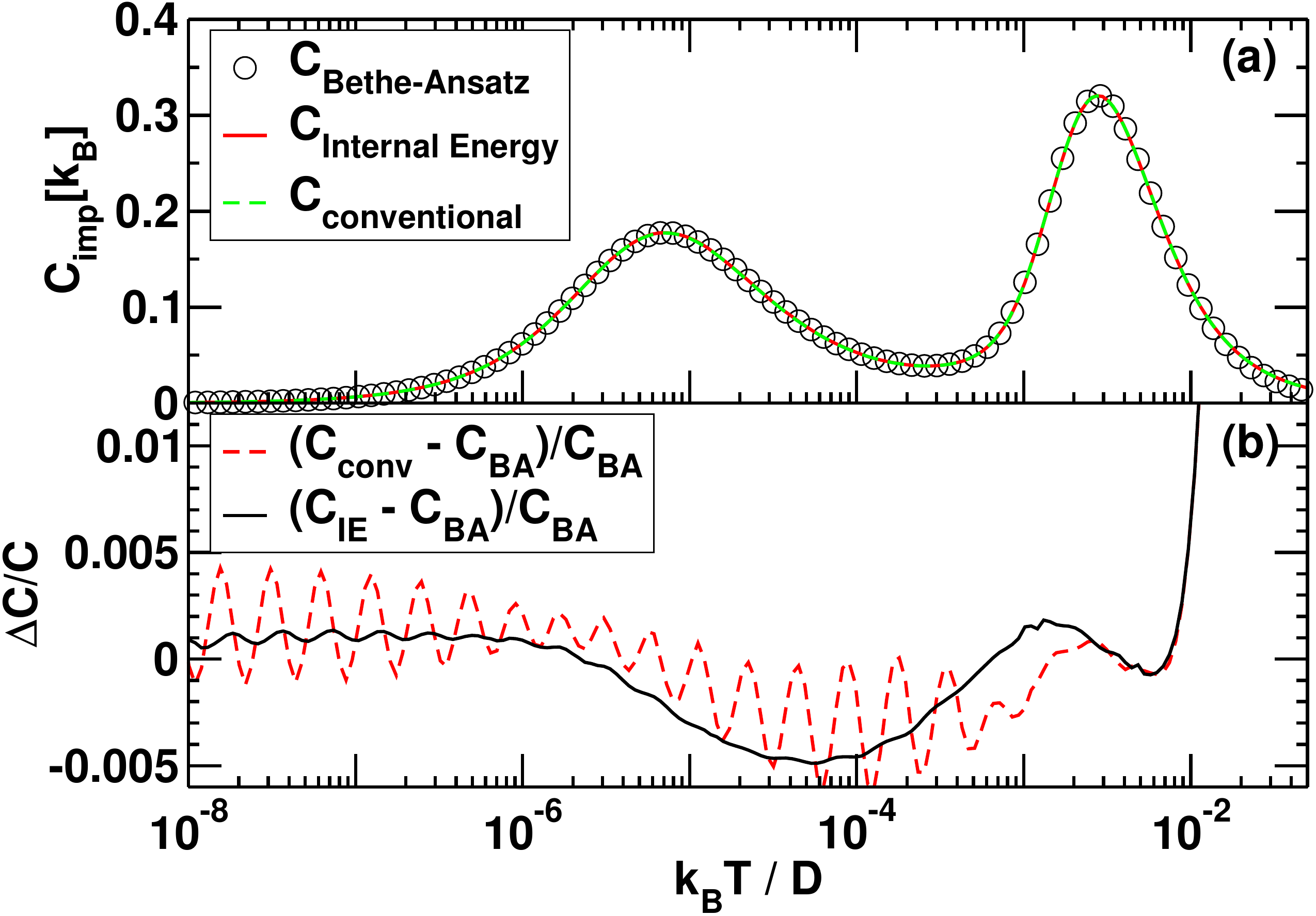}
\caption
{
  {\em (Color online)} 
  (a) Comparison of specific heat, $C_{\rm imp}(T)$, from the impurity internal energy (solid line) and 
  conventional approach (dashed line)  with the Bethe ansatz calculation  (symbols) for the symmetric
  Anderson model with parameters as in Fig.~\ref{fig1}. NRG parameters also as in Fig.~\ref{fig1} 
  with $n_{z}=2$.
  (b) The relative deviation with respect to the Bethe ansatz result of the new (solid line) and 
  conventional (dashed line) approaches.
  \label{fig6}
}
\end{figure}

\begin{figure}
\includegraphics[width=\linewidth,clip]{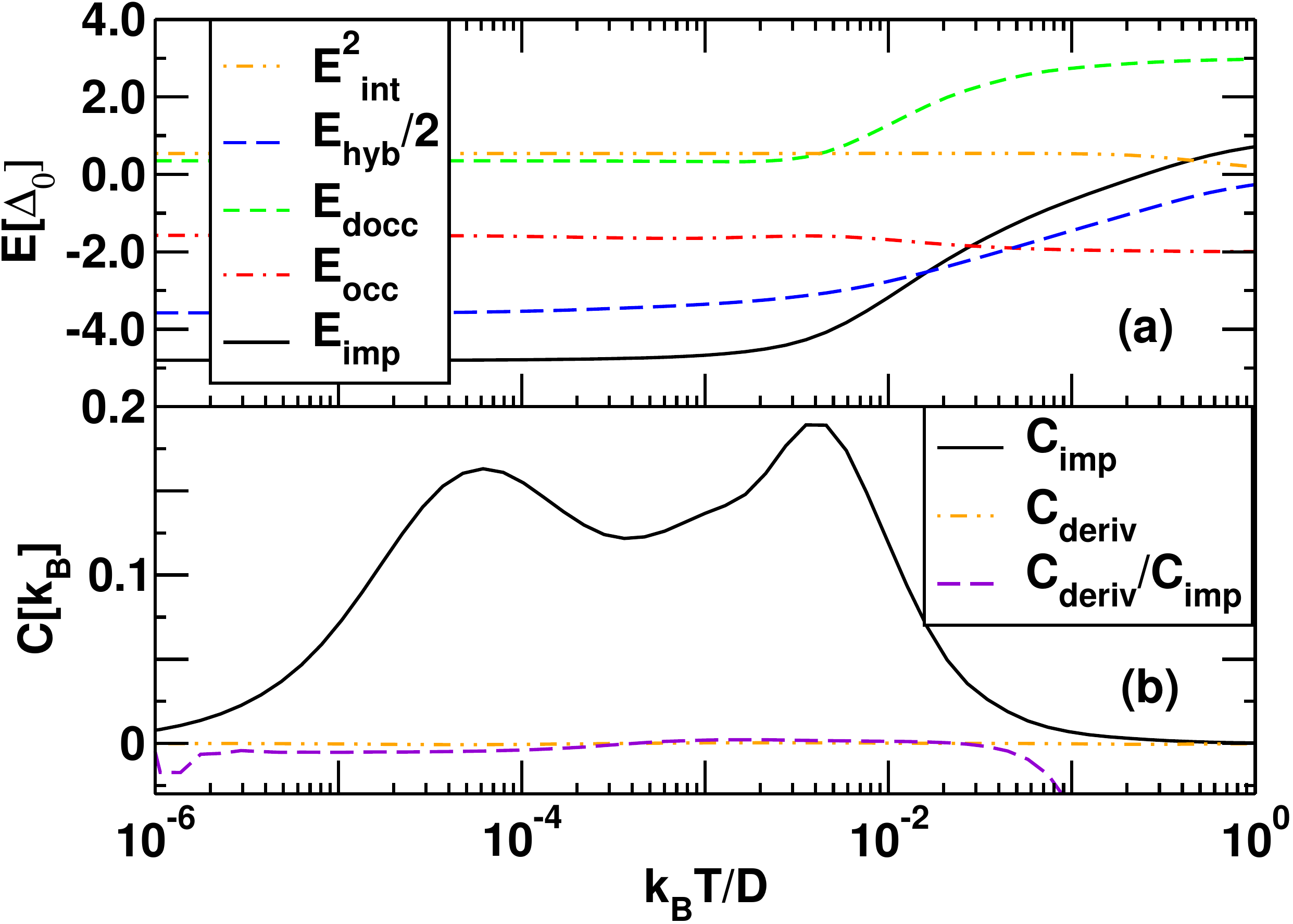}
\caption
{
  {\em (Color online)} 
  (a) The contribution $E_{\rm int}^{(2)}$ to $E_{\rm imp}$ as a function of temperature
  compared with $E_{\rm occ}$, $E_{\rm docc}$ and $E_{\rm hyb}$
  (in units of $\Delta_{0}$) for the asymmetric model. Model parameters: $U=12\Delta_{0}$, 
  $\Delta_{0}=0.001D$, $\varepsilon_{d}/\Delta_{0}=-2$ with a semi-elliptic
  hybridization function ${\rm Im} [\Delta(\omega)] 
  = -\frac{\Delta_{0}}{D}\sqrt{(D^2 -\omega^2)}$.  
  A small $\Lambda=1.5$ was used, which allows the spectral function entering $E_{\rm int}^{(2)}$ 
  to be obtained without $z$-averaging.
  (b) The contribution $C_{\rm deriv}=\partial E_{\rm int}^{(2)}(T)/\partial T$  to $C_{\rm imp}$.
  The relative size of $C_{\rm deriv}$ to $C_{\rm imp}$ lies between $0.2\%$ and $0.5\%$ for 
  all temperatures, except at temperatures approaching the bandwidth $D=1$.
  \label{fig7}
}
\end{figure}

We now quantify the error in neglecting $\partial E_{\rm int}^{(2)}(T)/\partial T$ in
Eq.~(\ref{eq:impurity-spec-from-energy}) for the calculation of impurity specific 
heats by, (a), comparing the result for $C_{\rm imp}$ obtained within
the new method with the Bethe ansatz calculations, and, (b), explicitly calculating
the contribution $\partial E_{\rm int}^{(2)}(T)/\partial T$. Figure~\ref{fig6}(a) shows the 
comparison to the Bethe ansatz calculation, where
we also include the specific heat from the conventional approach. The relative deviation of the NRG
calculations to the Bethe ansatz, shown in Fig.~\ref{fig6}(b), is below $1\%$ for all temperatures
$T< 0.01=10\Delta_{0}$. For $T\ll T_{\rm K}$, the relative error in $C_{\rm imp}$ from the 
internal energy is $0.1\%$ and $0.5\%$ in the conventional approach. The relative error
exhibits remnants of the discretization oscillations, which are not completely eliminated
with $z$-averaging. Notice also that the errors in the two
NRG calculations have the same error 
(relative to the Bethe ansatz) in the high temperature limit, $T\gg \Delta_{0}$. 
Hence, the latter error is not due to neglect of 
$E_{\rm int}^{(2)}$ in Eq.~(\ref{eq:impurity-internal-energy-exact}). Instead, it reflects, (a), the
different high energy cut-off schemes in NRG and Bethe ansatz, and, (b), the finite size errors in the 
high energy excitation spectrum in NRG, since the latter stem from the shortest chains diagonalized
(typically $m=4-6$), which are also the ones most sensitive to the logarithmic discretization. 
The fact that the errors in both NRG calculations also correlate at lower temperatures 
($T\lesssim \Delta_{0}$) suggests that the neglect of 
$E_{\rm int}^{(2)}$ in Eq.~(\ref{eq:impurity-internal-energy-exact}) is not the main source
of error in calculating $C_{\rm imp}(T)$. An explicit calculation that illustrates this is
shown in Fig.~\ref{fig7}. As stated above, the value of $E_{\rm int}^{(2)}$ is of order
$\Delta_{0}/\pi$, however, one clearly sees in Fig.~\ref{fig7}(a) that $E_{\rm int}^{(2)}$
has little temperature dependence (relative to the other contributions) for all temperatures
extending up to the bandwidth $D=1$. It's relative contribution to the impurity specific
heat, shown in Fig.~\ref{fig7}(b), for an energy dependent $\Delta(\omega)$, 
is negligible, typically contributing below $0.5\%$.

\section{Results for the symmetric model} 
\label{sec:symmetric model results}
In this section we show results for the entropy and specific heat of the 
Anderson model at the particle-hole symmetric point $\varepsilon_{d}=-U/2$.
Results for zero magnetic field and increasing correlation strength 
$U/\Delta_{0}$ are presented in Sec.~(\ref{subsec:zero field}) and results 
for finite magnetic fields are given in Sec.~(\ref{subsec:finite field}).

The  symmetric Anderson model has been investigated in detail 
\cite{Hewson1997} and is well understood. For $U/\Delta_{0}\gg 1$ 
and $-\varepsilon_{d}\gg \Delta_{0}$, a local
spin $S=1/2$ magnetic moment forms on the impurity. In this limit, the
physics of the symmetric model at low temperatures 
$T\ll {\rm min}(|\varepsilon_{d}+U|,|\varepsilon_{d}|,D)$
is that of the Kondo model

\begin{equation}
\label{eq:KondoModel}
H_{\rm K} = H_{0} + J \vec{S}.\vec{s}_{0},
\end{equation}
where, $J$ is an antiferromagnetic exchange coupling between the
local spin $S$ and the conduction electron spin-density $\vec{s}_{0}$
at the impurity site. The value of $J$ is given by the Schrieffer-Wolff
transformation\cite{Schrieffer1966} $J=4V^{2}/U$. 
The low temperature properties (for $U\gg \Delta_{0}$) are universal functions
of $T/T_{\rm K}$ and $B/T_{\rm K}$ where we choose to define the Kondo scale from
the Bethe ansatz result for the $T=0$ susceptibility $\chi(0)$ via 
$\chi(0)=(g \mu_{\rm B})^{2}/4T_{\rm K}$. For $U\gg\Delta_{0}$, $T_{\rm K}$
is given by 
\begin{equation}
\label{eq:symmetric-TK}
T_{\rm K}=\sqrt{U\Delta_{0}/2}e^{-\pi U/8\Delta_{0} + \pi \Delta_{0}/2U},
\end{equation}
within corrections which are exponentially small in 
$U/\pi\Delta_{0}$ (see Ref.~\onlinecite{Hewson1997}). For $U=0$, the
symmetric Anderson model reduces to a resonant level model and the relevant
low temperature scale is then $\Delta_{0}$.

\subsection{Zero magnetic field} 
\label{subsec:zero field}
\begin{figure}
\includegraphics[width=\linewidth,clip]{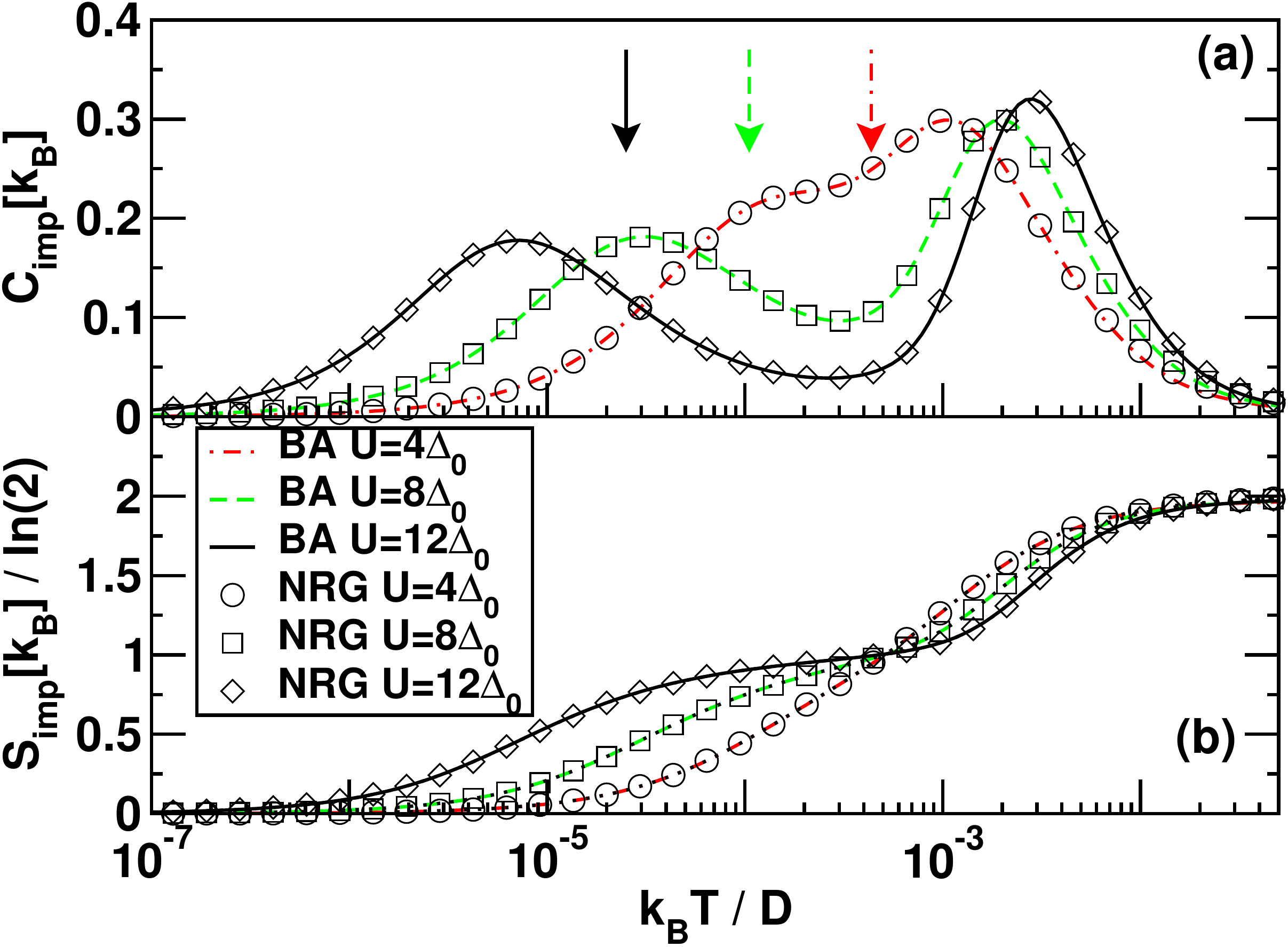}
\caption{{\em (Color online)} 
  Temperature dependence of, (a), the impurity specific heat, $C_{\rm imp}(T)$, 
  and, (b), the impurity entropy, $S_{\rm imp}(T)/\ln(2)$, for the symmetric Anderson
  model with $\Delta_{0}=0.001D$ and increasing values of the Coulomb interaction:  
  $U/\Delta_{0} = 4, 8, 12$. Arrows in (a) indicated the Kondo scale 
  $T_{\rm K}$ defined in Eq.~\ref{eq:symmetric-TK}. 
  Symbols: new approach using NRG with $\Lambda = 4$ with an  energy 
  cut-off $e_{c}(\Lambda=4)=40$, and $z$-averaging [$n_{z}=2$, $z=1/4,\, 3/4$].
  Lines: corresponding Bethe ansatz calculations.
\label{fig8}}
\end{figure}

A comparison of the new approach with Bethe ansatz calculations is shown in
Fig.~\ref{fig8} for the temperature dependence of the impurity specific 
heat and entropy for increasing values of the Coulomb interaction
$U/\Delta_{0}$. For $U/\Delta_{0}=12$, the Kondo induced peak in the specific heat
at $T_{\rm p}= \alpha T_{\rm K}$ with $\alpha\approx 0.29$ is well separated from 
the peak at $T\approx |\varepsilon_{d}|$ due to the resonant level. With
decreasing $U/\Delta_{0}$, the Kondo effect is suppressed and the Kondo induced
peak in $C_{\rm}(T)$ eventually merges with the peak due to the resonant level for 
$U/\Delta_{0}\rightarrow 0$. Good agreement between the NRG and the exact
Bethe ansatz calculations is seen for all values of $U/\Delta_{0}$.

\subsection{Finite magnetic field} 
\label{subsec:finite field}
At finite magnetic fields $B>0$, the ${\rm SU}(2)$ spin symmetry which we use
in the NRG calculations, is broken. Therefore, in order to carry out 
calculations at finite magnetic field $B>0$, 
preserving the numerical advantages of the full ${\rm SU}(2)$ symmetry, such as the
increased number of states that can be retained,
we obtained the finite field results by mapping the symmetric positive $U$ Anderson
model onto the negative-$U$ Anderson model in the absence of a magnetic field but with
local level given by $\varepsilon_{d}= -U/2 - B/2$ with $U$ negative. \cite{Iche1972,Hewson2006}
This correspondence results from a particle-hole transformation on the
down spins only: $d{_\downarrow}\rightarrow d_{\downarrow}^{\dagger},\, 
d{_\uparrow}\rightarrow d_{\uparrow}$, and 
$c_{k\downarrow}\rightarrow c_{-k\downarrow}^{\dagger}, 
c_{k\uparrow}\rightarrow c_{k\uparrow}$ with a particle-hole symmetric band
$\epsilon_{k}=-\epsilon_{-k}$.

\begin{figure}
\includegraphics[width=\linewidth,clip]{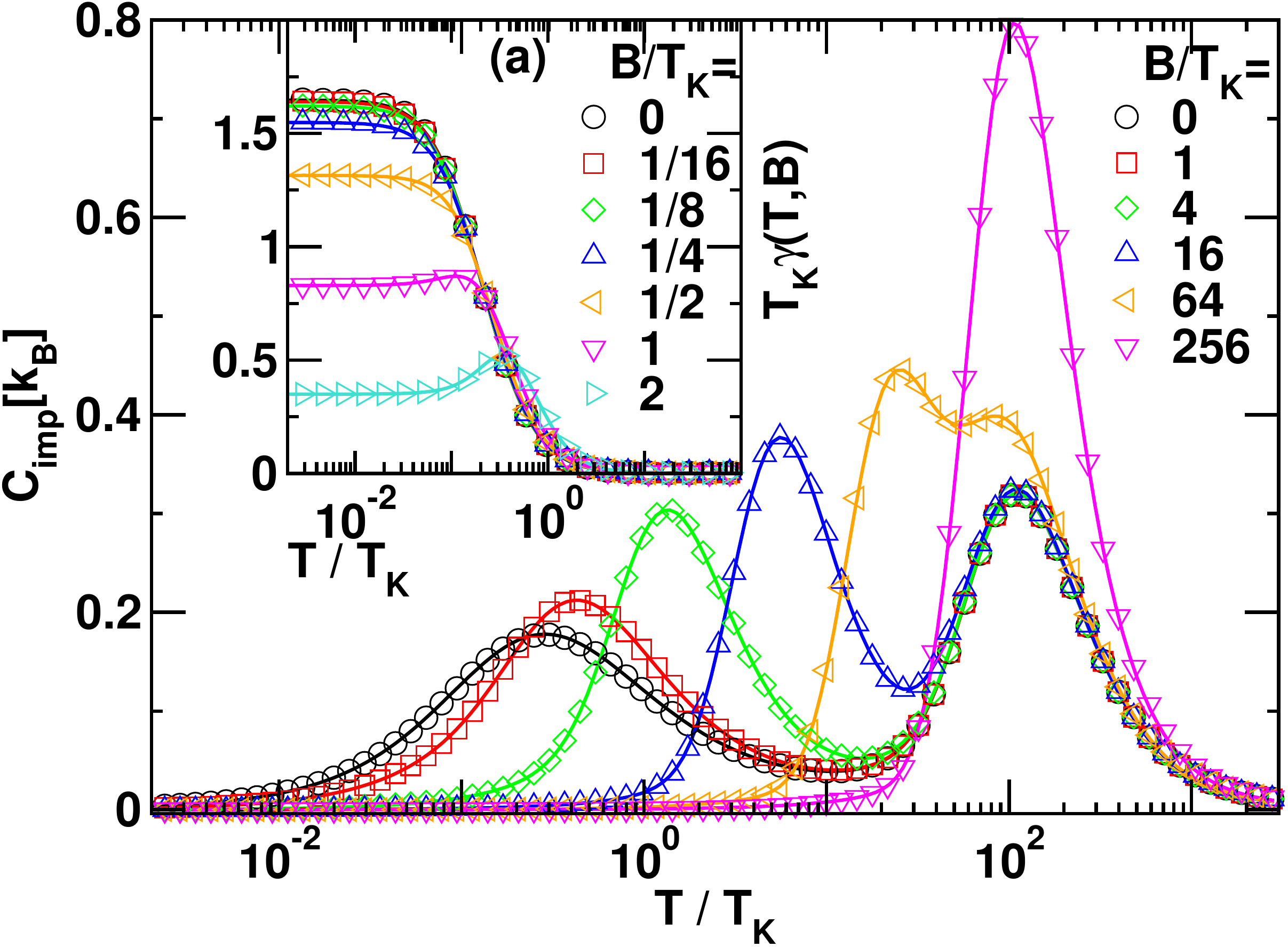}
\caption{{\em (Color online)}
  Temperature dependence of the impurity specific heat, $C_{\rm imp}(T,B)$, 
  for the symmetric Anderson model for $U/\Delta_{0} = 12$, $\Delta_{0}=0.001D$ and 
  increasing values of the magnetic field $B/T_{\rm K}\ge 1$ where the 
  Kondo scale $T_{\rm}$ is defined in Eq.~\ref{eq:symmetric-TK}.
  Symbols: NRG calculations $\Lambda = 4$ with an  energy 
  cut-off $e_{c}(\Lambda=4)=40$, and $z$-averaging [$n_{z}=2$, $z=1/4,\, 3/4$].
  Lines: Bethe ansatz calculations.
  Inset (a): $T_{\rm K}\gamma(T,B)$ versus $T/T_{\rm K}$ for several 
  values of $B/T_{\rm K}\le 2$, where $\gamma(T,B) = C_{\rm imp}(T,B)/T$. 
\label{fig9}}
\end{figure}

Figure~\ref{fig9} shows the temperature dependence of $C_{\rm imp}(T,B)$ 
for $B/T_{\rm K}\ge 1$ using our new approach and compared with Bethe ansatz 
calculations. The Kondo peak in the specific heat shifts to higher fields
with increasing $B$ and its position scales as $B^{2}/T_{\rm K}$ for 
$T_{\rm K}\ll B \ll \varepsilon_{d}$. In contrast, the resonant level peak remains approximately
fixed at $T\approx \varepsilon_{d}$. As $B$ approaches the value $\varepsilon_{d}$,
the two peaks merge into one peak at $T\approx \varepsilon_{d}$, with aproximately
twice the height of the $B=0$ resonant level peak, and containing the whole
entropy $S_{\rm imp}/k_{\rm B}=\ln(4)$. The low field behaviour of $C_{\rm imp}(T,B)$, 
also compared to Bethe ansatz calculations, is shown in Fig.~\ref{fig9}(a) as 
$T_{\rm K}\gamma(T,B)=C_{\rm imp}(T,B)/(T/T_{\rm K})$ versus $T/T_{\rm K}$ 
for $B/T_{\rm K}\le 2$. For $T,B\rightarrow 0$, $\gamma(T,B)\rightarrow \gamma(0,0)
\sim 1/T_{\rm K}$ where $\gamma(0,0)$ is the linear coefficient of specific heat. 
This is strongly enhanced for $U/\Delta_{0}\gg 1$ due to the exponential decrease of 
$T_{\rm K}$. A finite magnetic field of order $T_{\rm K}$ significantly 
suppresses the Kondo effect and results in smaller values of $\gamma(0,B)$.
As another check on the accuracy of our calculations, we estimate the
Wilson ratio $R_{\rm W}=4\pi^{2}\chi(0)/3\gamma(0,0)$. This takes the value $2$ in
the Kondo regime of the symmetric Kondo model (i.e., for $U\gg \Delta_{0}$)
From the definition of $T_{\rm K}$, we have that the susceptibility 
$\chi(0)=1/4T_{\rm K}$, and from Fig.~\ref{fig9}(a) we extract 
$\gamma(0,0)\approx 1.64/T_{\rm K}$, resulting in $R_{\rm W}\approx 2.006$, that is,
a relative error in $R_{\rm W}$ below $1\%$.

\section{Results for the asymmetric model} 
\label{sec:asymmetric model results}
\begin{figure}
\includegraphics[width=\linewidth,clip]{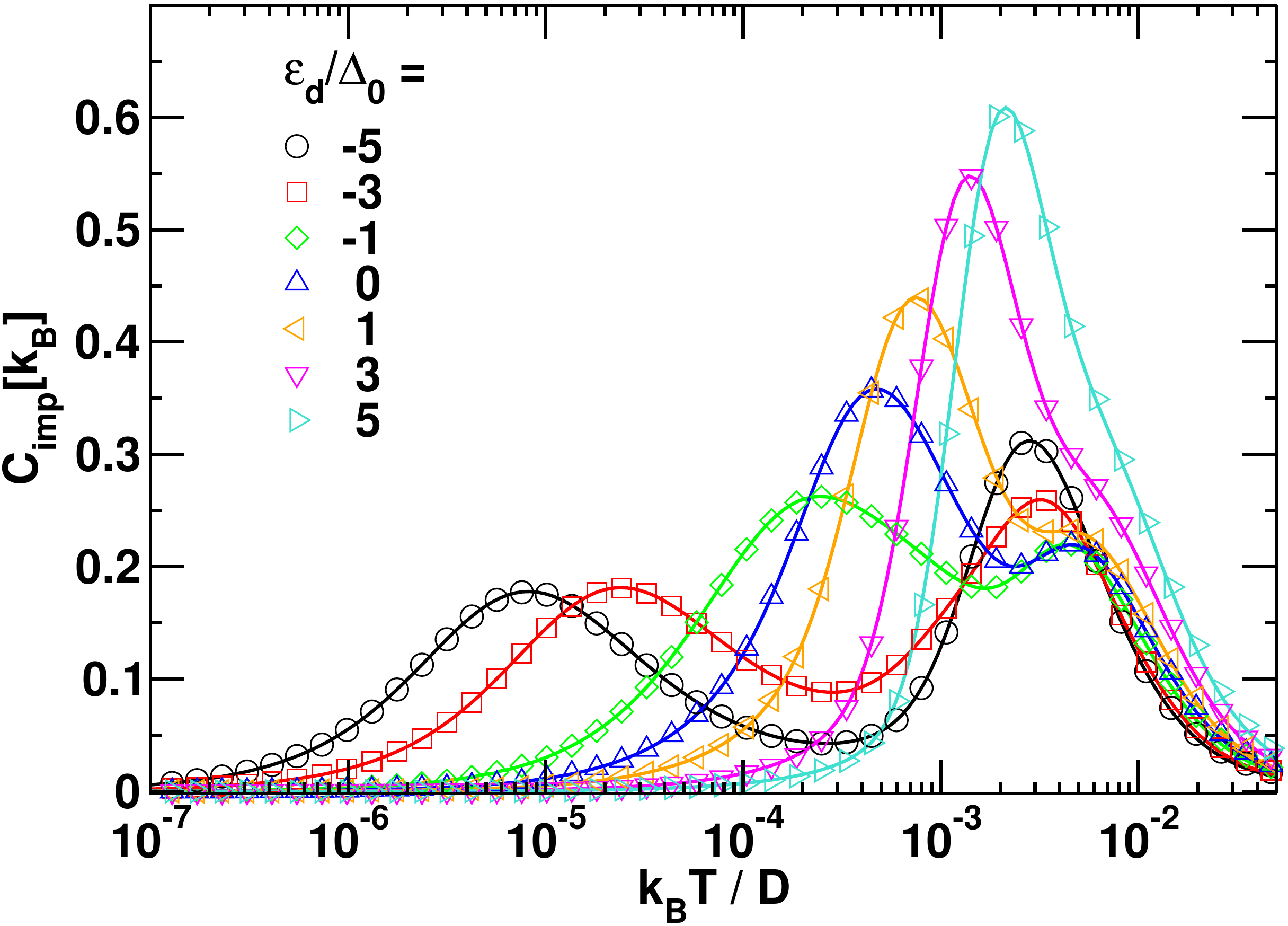}
\caption{{\em (Color online)} 
Temperature dependence of $C_{\rm imp}(T)$ for $U/\Delta_{0} = 12,\,\Delta_{0}=0.001D$ and local
level positions $\varepsilon_{d}/\Delta_{0}$ ranging from Kondo ($\varepsilon_{d}/\Delta_{0} 
\le -1$), mixed valence ($|\varepsilon_{d}/\Delta_{0}|\le 1$), and empty orbital
($\varepsilon_{d}/\Delta_{0}> 1$) regimes. Symbols: NRG calculations (new approach, $z$-averaging and NRG
parameters as in Fig.~\ref{fig1}). 
Lines: Bethe ansatz calculations.
  \label{fig10}}
\end{figure}
Figure~\ref{fig10} shows the impurity specific heat versus temperature
for the asymmetric Anderson model, that is, for $\varepsilon_{d}>-U/2$,
calculated within the new approach. For comparison, we also show the corresponding Bethe
ansatz calculations. One sees again excellent agreement at all temperatures between
the two methods. Results for $\varepsilon_{d}<-U/2$ are not shown, since these can
be obtained from results for $\varepsilon_{d}>-U/2$ by noting that the Anderson model with parameters 
$\varepsilon_{d},U,V$ transforms, under a particle-hole transformation applied to
both spin species, to an Anderson model with parameters $-(\varepsilon_{d}+U),U,V$.
This holds for a particle-hole symmetric constant density of states,  
the case considered here.

\begin{figure}
\includegraphics[width=\linewidth,clip]{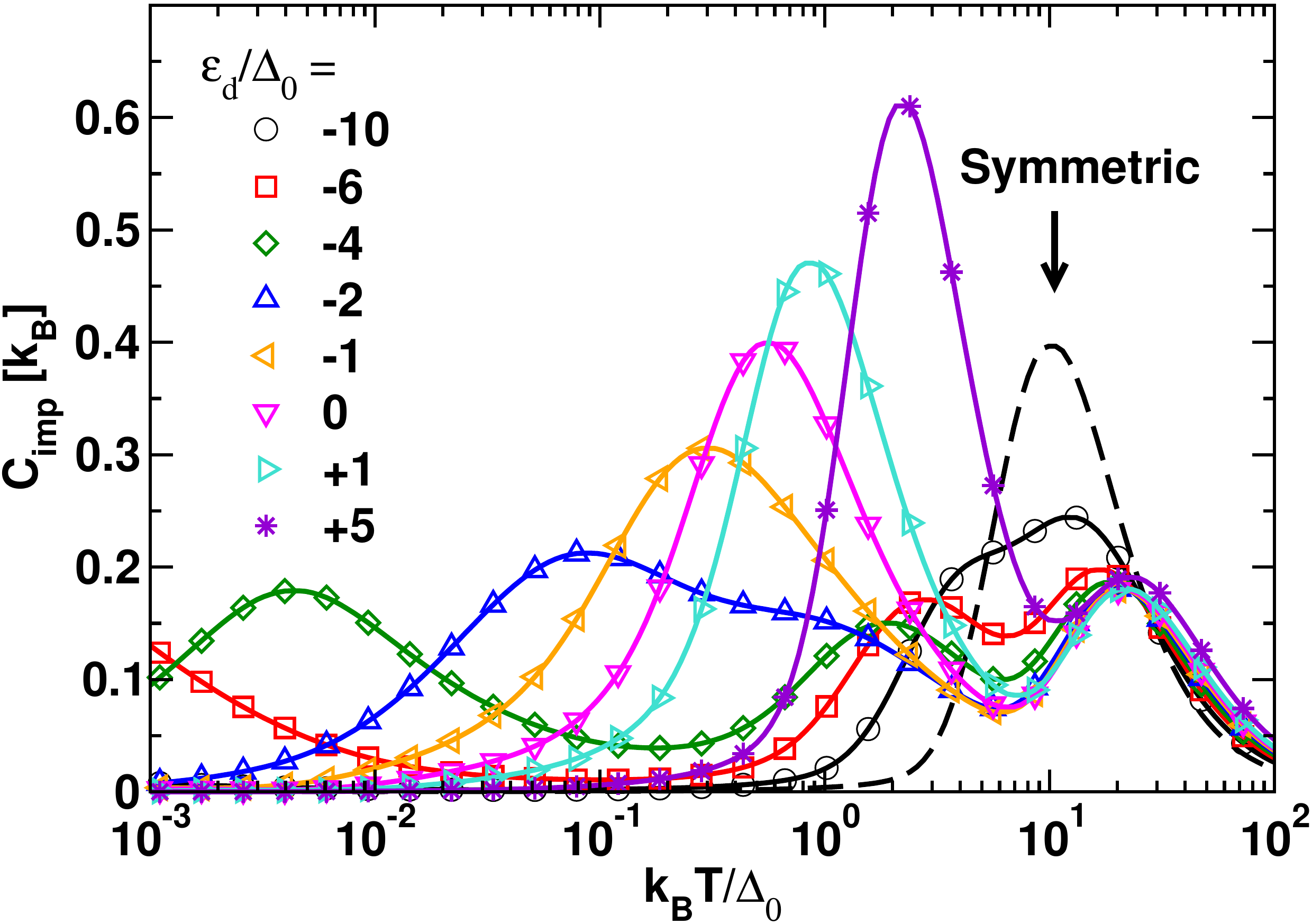}
\caption{{\em (Color online)} 
Temperature dependence of $C_{\rm imp}(T)$ for $U/\Delta_{0} = 48,\,\Delta_{0}=0.0001D$ 
and local level positions $\varepsilon_{d}/\Delta_{0} = -10, -6, -4, -2$ (Kondo regime), 
$\varepsilon_{d}/\Delta_{0} =-1,0,+1$ (mixed valence regime) and $\varepsilon_{d}/\Delta_{0} = +5$
(empty orbital regime). NRG using the new approach (symbols) and conventional approach (solid lines) 
[$\Lambda=20, n_{z}=4, e_{c}(\Lambda)=130$]. Dashed line: 
resonant level peak in $C_{\rm imp}$ at $T/\Delta_{0}\approx |\varepsilon_{d}|/\Delta_{0}=U/2\Delta_{0}=24$ 
for the symmetric model (the Kondo induced peak at much lower $T$ is not shown). The two
high temperature peaks of the asymmetric model evolve from this peak when the asymmetry is finite.
  \label{fig11}}
\end{figure}

The specific heat curves for the asymmetric model are more complicated than those of the
symmetric model. In the latter, the relevant excitations were the low temperature spin flip
excitations, characterized by the Kondo scale $T_{\rm K}$, and the excitations involving
addition or removal of an electron from the resonant level, both characterized by 
an energy $|\varepsilon_{d}|=U/2$.
This accounts for the two peaks in the specific heat of the symmetric model: A high temperature 
peak at $T\approx |\varepsilon_{d}|$ and a low temperature Kondo induced peak at $T\approx T_{\rm K}$. 
For the asymmetric Anderson model, three types of excitation are possible: Low-temperature
spin flip excitations, associated with the Kondo scale $T_{\rm L}= \sqrt{U\Delta_{0}/2}
e^{-\pi|\varepsilon_{d}|\,|\varepsilon_{d}+U|/2U\Delta_{0}}$ of the
asymmetric model,\cite{Hewson1997} and excitations associated with, (i), removing 
an electron from a singly occupied level (with energy scale $|\varepsilon_{d}|$) and (ii), 
removing an electron from a doubly occupied level (with energy scale $|\varepsilon_{d}+U|$).
Thus, three peaks can be present in $C_{\rm imp}(T)$: a Kondo induced peak at $T\approx T_{\rm L}$, 
and two charge fluctuation induced peaks at $T\approx T_{1}=|\varepsilon_{d}|$ 
and $T\approx T_{2}=|\varepsilon_{d}+U|$, respectively. 
In Fig.~\ref{fig10}, the two high temperature peaks are seen in the mixed valence regime
and partly also in the empty orbital regime (where the upper peak at $T_{2}$ appears 
as a shoulder of the main peak at $T_{1}$). However, in the Kondo regime, the cases
$\varepsilon_{d}/\Delta_{0}=-5,-3$ with the choice $U=12\Delta_{0}$ result in $T_{1}/\Delta_{0}=5,3$
and $T_{2}/\Delta_{0}=7,9$. In these cases, $T_{1}$ and $T_{2}$ are too close for 
separate peaks to be seen. In order to clarify this, we carried out calculations
for $U=48\Delta_{0}\gg \Delta_{0}$, and $\varepsilon_{d}/\Delta_{0}=-10,-8,-6,-4,-2$ in the
Kondo regime, for which
$T_{1}/\Delta_{0}=10,8,6,4,2$ and $T_{2}/\Delta_{0}=38,40,42,44,46\gg T_{1}/\Delta_{0}$ 
are disparate scales. Figure~\ref{fig11} shows
how the peaks at $T\approx T_{1}$ and $T\approx T_{2}$ evolve from the peak at 
$T\approx |\varepsilon_{d}|=U/2$ of the symmetric model (dashed line in Fig.~\ref{fig11}) 
on increasing $\varepsilon_{d}$ above $-U/2$. Simultaneously, the Kondo peak in the 
specific heat at $T_{\rm L}$ shifts to higher temperatures and eventually merges with
the peak at $T_{1}$ when the mixed valence regime is reached (i.e., for $\varepsilon_{d}=-\Delta_{0}$).
Thereafter, only the high temperature peaks at $T_{1}$ and $T_{2}$ are present. Notice also,
that in the mixed valence regime $T_{1}$ differs significantly from $|\varepsilon_{d}|$, a
result of non-trivial renormalizations present in the mixed valence regime, but absent in the
empty orbital regime.

\section{Generalization to other models}
\label{sec:other models}
The approach of Sec.~\ref{sec:new approach} can be straightforwardly 
generalized to multiorbital and multichannel Anderson impurity models with 
arbitrary local Coulomb interactions, as we briefly outline in Sec.~\ref{sec:multichannel}.
In addition, in Sec.~\ref{sec:dissipative two state} we discuss it's application to 
dissipative two state systems and the anisotropic Kondo model (AKM). 

\subsection{Multiorbital and multichannel Anderson models}
\label{sec:multichannel}
The multiorbital and multichannel Anderson impurity model
is given by $H=H_{\rm imp}+ H_{\rm 0} +H_{\rm int}$, where 
$H_{\rm imp}  = \sum_{\alpha\sigma}\varepsilon_{\alpha}d_{\alpha\sigma}^{\dagger}d_{\alpha\sigma} + H_{\rm C}(U,U',J)$, describes the impurity with a set of local levels 
having energies $\varepsilon_{d\alpha}, \alpha=1,\dots,g$
and $H_{\rm C}(U,U',J)$ is the local Coulomb interaction involving intra-orbital 
$U$, inter-orbital $U'$ and a Hund's exchange term $J$. The conduction electrons 
are described by
 $H_{\rm 0} =  \sum_{k\alpha\sigma}\epsilon_{k\alpha}c_{k\alpha\sigma}^{\dagger}c_{k\alpha\sigma}$
where $\epsilon_{k\alpha}$ is the kinetic energy of electrons in band $\alpha$.
These bands hybridize with hybridization strengths $V_{\alpha},\alpha=1,\dots,g$ 
to the local levels via
 $H_{\rm int}  =  \sum_{k\alpha\sigma}V_{\alpha}
(c_{k\alpha\sigma}^{\dagger}d_{\alpha\sigma} +
d^{\dagger}_{\alpha\sigma}c_{k\alpha\sigma})$. 
Let $\Delta_{\alpha}(\omega)=\sum_{k}V_{\alpha}^{2}/(\omega-\epsilon_{k\alpha})$ denote the
hybridization functions characterizing $H_{\rm int}$. Proceeding as in Sec.~\ref{sec:new approach},
we write the impurity internal energy as $E_{\rm imp}=E_{\rm total}-E_{\rm 0}$ where
$E_{\rm total}=\langle H\rangle$ is the total energy and 
$E_{\rm 0}= \langle H_{0}\rangle=\sum_{k\alpha\sigma}\epsilon_{k\alpha}\langle 
c_{k\alpha\sigma}^{\dagger}c_{k\alpha\sigma}\rangle_{\rm 0}$ is the energy of the non-interacting
conduction electrons in the absence of the impurity. The latter is given by
$
E_{0} = \sum_{\alpha\sigma}\int d\epsilon f(\epsilon)\epsilon N_{\alpha}(\epsilon),
$
where $f(\epsilon)$ is the Fermi function and $N_{\alpha}(\epsilon)=\sum_{k}\delta(\epsilon-\epsilon_{k\alpha})$ is the non-interacting conduction electron density of states per spin for band $\alpha$.
$E_{\rm total}$ is a sum of local occupation number contributions 
$E_{\rm occ}=\sum_{\alpha\sigma}\varepsilon_{\alpha}\langle n_{\alpha\sigma}\rangle$ and local Coulomb
terms $E_{C}=\langle H_{\rm C}(U,U',J)\rangle$ and two further terms 
involving the interacting band 
$E_{\rm cond}=\sum_{k\alpha\sigma}\epsilon_{k\alpha}\langle 
c_{k\alpha\sigma}^{\dagger}c_{k\alpha\sigma}\rangle$ 
and the hybridization energy $E_{\rm hyb}= \sum_{\alpha\sigma}V_{\alpha}\langle
d_{\alpha\sigma}^{\dagger}f_{0\alpha\sigma} + H.c.\rangle$ where 
$V_{\alpha}f_{0\alpha\sigma}=\sum_{k}c_{k\alpha\sigma}$:
\begin{equation}
E_{\rm total} = E_{\rm occ} + E_{\rm C} + E_{\rm cond} + E_{\rm hyb}.
\end{equation}
We evaluate the latter two contributions as in Sec.~\ref{sec:new approach}, finding
\begin{equation}
E_{\rm hyb} = -\frac{2}{\pi}\sum_{\alpha\sigma}\int d\omega f(\omega) {\rm Im} \left[
G_{d\alpha\sigma}(\omega)\Delta_{\alpha}(\omega)
\right],
\label{hyb-dynamics-multi}
\end{equation}
and $E_{\rm cond}=E_{\rm 0} + E_{\rm int}$, where
\begin{eqnarray}
E_{\rm int} &=& \frac{1}{\pi}\sum_{\alpha\sigma}\int d\omega f(\omega){\rm Im}
\left[G_{d\alpha\sigma}(\omega)
\frac{\partial}{\partial\omega}\left(\omega\Delta_{\alpha}(\omega)\right)
\right]\nonumber\\
&=& E_{\rm int}^{(1)} + E_{\rm int}^{(2)},\label{band-contribution-multi}\\
E_{\rm int}^{(1)} &=&\frac{1}{\pi}\sum_{\alpha\sigma}\int d\omega f(\omega){\rm Im}
\left[G_{d\alpha\sigma}(\omega)\Delta_{\alpha}(\omega)\right]\\
E_{\rm int}^{(2)} &=&\frac{1}{\pi}\sum_{\alpha\sigma}\int d\omega f(\omega){\rm Im}
\left[G_{d\alpha\sigma}(\omega)\omega\frac{\partial\Delta_{\alpha}(\omega)}{\partial\omega}\right],
\end{eqnarray}
and $G_{d\alpha\sigma}(\omega)$ is the retarded Green function for local level $\alpha$.
Combining $E_{\rm int}^{(1)}$ with $E_{\rm hyb}$ gives for the impurity internal energy
\begin{equation}
\label{eq:impurity-internal-energy-exact-multi}
E_{\rm imp} = E_{\rm occ} + E_{\rm C} + \frac{1}{2}E_{\rm hyb} + E_{\rm int}^{(2)},
\end{equation}
where, as before, all contributions except the last one are evaluated as local static
correlation functions. For reasons discussed in Sec.~\ref{sec:new approach}, the temperature
dependence of the last term is negligible in many cases and the impurity specific heat
can be calculated to high accuracy via
\begin{eqnarray}
\label{eq:impurity-spec-from-energy-multi}
C_{\rm imp} &=& \frac{\partial E_{\rm occ}}{\partial T} + \frac{\partial E_{\rm C}}{\partial T} 
+ \frac{1}{2}\frac{\partial E_{\rm hyb}}{\partial T}\\
&=& \frac{\partial E_{\rm ionic}}{\partial T} + \frac{1}{2}\frac{\partial E_{\rm hyb}}{\partial T},
\end{eqnarray}
where $E_{\rm ionic}=\langle H_{\rm imp}\rangle$.

\subsection{Dissipative two state systems and the anisotropic Kondo model}
\label{sec:dissipative two state}
The method of Sec.~\ref{sec:new approach} can be applied to bosonic models such as the 
dissipative two state system,\cite{Leggett1987,Weiss2008} and for Ohmic dissipation, one
can further relate the results to the AKM and related models (for example, a two-level system
in a metallic environment \cite{Ramos2003}). Dissipative two state
systems are of interest in many contexts, including the description of qubits 
coupled to their environment. 

The Hamiltonian of the dissipative two state system is given by $H=H_{\rm S}+H_{\rm B} +H_{\rm I}$.
The first term $H_{\rm S}=-\frac{1}{2}\Delta_{0}\sigma_{x}+\frac{1}{2}\epsilon\sigma_{z}$ describes
a two-level system with bias splitting $\epsilon$ and tunneling amplitude $\Delta_{0}$, and
$\sigma_{i=x,y,z}$ are Pauli spin matrices.
$H_{\rm B}=\sum_{i}\omega_{i}(a^{\dagger}_{i}a_{i}+1/2)$ is the environment and consists of
an infinite set of harmonic oscillators ($i=1,2,\dots,\infty$) with $a_{i} (a_{i}^{\dagger})$ the
annihilation (creation) operators for a harmonic oscillator of frequency $\omega_{i}$ and 
$0\le \omega_{i}\le \omega_{\rm c}$, where $\omega_{\rm c}$ is an upper cut-off frequency. The non-interacting
density of states of the environment is denoted by $g(\omega_{i})=\sum_{i}\delta(\omega-\omega_{i})$ and
is finite in the interval $[0,\omega_{\rm c}]$ and zero otherwise.
Finally, $H_{\rm I}=\frac{1}{2}\sigma_{z}\sum_{i}\lambda_{i}(a_{i}+a_{i}^{\dagger})$ describes
the coupling of the two-state system co-ordinate $\sigma_{z}$ to the oscillators, with
$\lambda_{i}$ denoting the coupling strength to oscillator $i$. The function 
$\Gamma(\omega+i\delta)=\sum_{i}(\lambda_{i}/2)^2/(\omega-\omega_{i}+i\delta)=
\int d\omega' (\lambda(\omega')/2)^2\,g(\omega')/(\omega-\omega'+i\delta)$ 
characterizes the system-environment interaction. The Ohmic two state system, 
specified by a spectral function $J(\omega)=-\frac{1}{\pi}{\rm Im}\Gamma(\omega+i\delta)
\sim \alpha\omega$ for $\omega\rightarrow 0$, where $\alpha$ is the dimensionless dissipation
strength, is equivalent to the AKM $H=\sum_{k\sigma}\epsilon_{k}c_{k\sigma}^{\dagger}c_{k\sigma}
+\frac{J_{\perp}}{2}(S^{+}s_{\rm 0}^{-}+S^{-}s_{\rm 0}^{+})+J_{\parallel}S_{z}s_{0}^{z}+BS_{z}$,
where $J_{\perp}$ ($J_{\parallel}$ is the transverse (longitudinal) part of the Kondo exchange
interaction and $B$ is a local magnetic field.
The correspondence is given by $\rho J_{\perp}=-\Delta_{0}/\omega_{\rm c}$
and $\alpha = (1+2\delta/\pi)^2$ where $\delta=\arctan(-\pi\rho J_{\parallel}/4)$ and
$\rho$ is the density of states of the conduction electrons in the AKM.
\cite{Hakim1985,Costi1996,Leggett1987,Costi1999,Weiss2008} The low energy
scale of the Ohmic two state system is the renormalized tunneling amplitude $\Delta_{\rm r}$
given by $\Delta_{\rm r}/\omega_{\rm c}=(\Delta_{\rm 0}/\omega_{\rm c})^{1/(1-\alpha)}$ and 
corresponds to the low energy Kondo scale $T_{\rm K}$ of the AKM. 
Special care is needed to obtain results for the Ohmic two state system from the AKM
in the vicinity of the singular point $\alpha\rightarrow 1^{-}$, since this corresponds to 
$J_{\parallel} \rightarrow 0$ but with the condition $0<J_{\perp}<J_{\parallel}$, 
that is, in terms of parameters of the Ohmic two state system one requires 
$\Delta_{\rm 0}/\omega_{\rm c}\ll 1-\alpha \ll 1$ in order to investigate the 
vicinity of $\alpha=1$ within the AKM.\cite{Weiss2008} 

\begin{figure}
\includegraphics[width=\linewidth,clip]{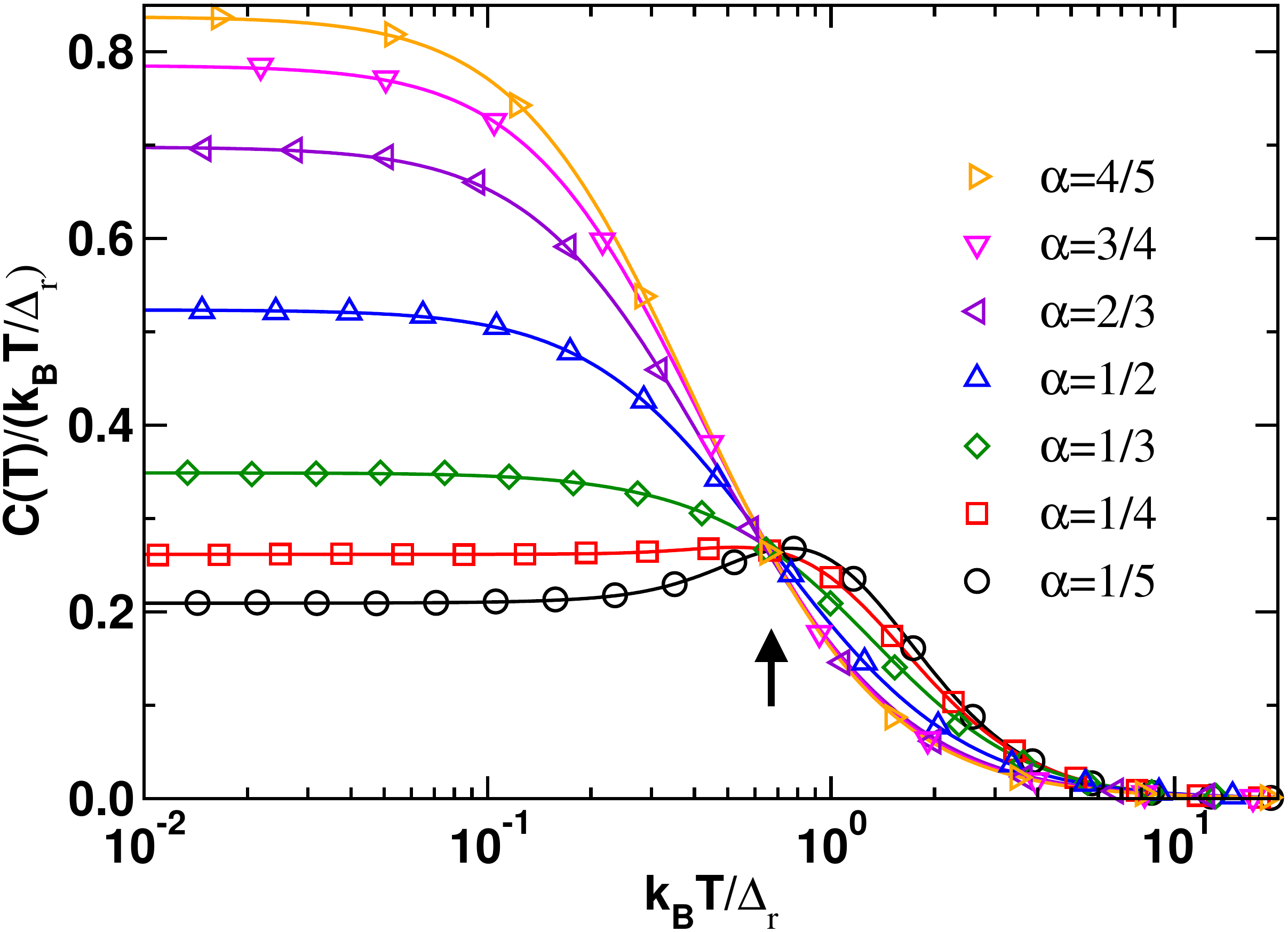}
\caption{{\em (Color online)} 
  Specific heat, $C_{\rm imp}(T)/{k_{\rm B}T/\Delta_{\rm r}}$, of the Ohmic 
  two state system as a function of reduced
  temperature $k_{B}T/\Delta_{\rm r}$, for a range of dissipation strengths 
  $\alpha=1/5,1/4,1/3,1/2,2/3,3/4,4/5$. Symbols: NRG results in new approach. Lines: Bethe ansatz results. 
  The renormalized tunneling amplitude $\Delta_{\rm r}$ from the Bethe ansatz is used.
  The vertical arrow indicates 
  the approximate crossing point at $k_{\rm B}T/\Delta_{\rm r}\approx 0.67$.
  Model parameters: $\Delta_{\rm 0}/\omega_{\rm c}=0.005$. NRG parameters:
  $\Lambda = 10, n_{z}=4$ retaining $860$ states per NRG iteration.
\label{fig12}}
\end{figure}

\begin{figure}
\includegraphics[width=\linewidth,clip]{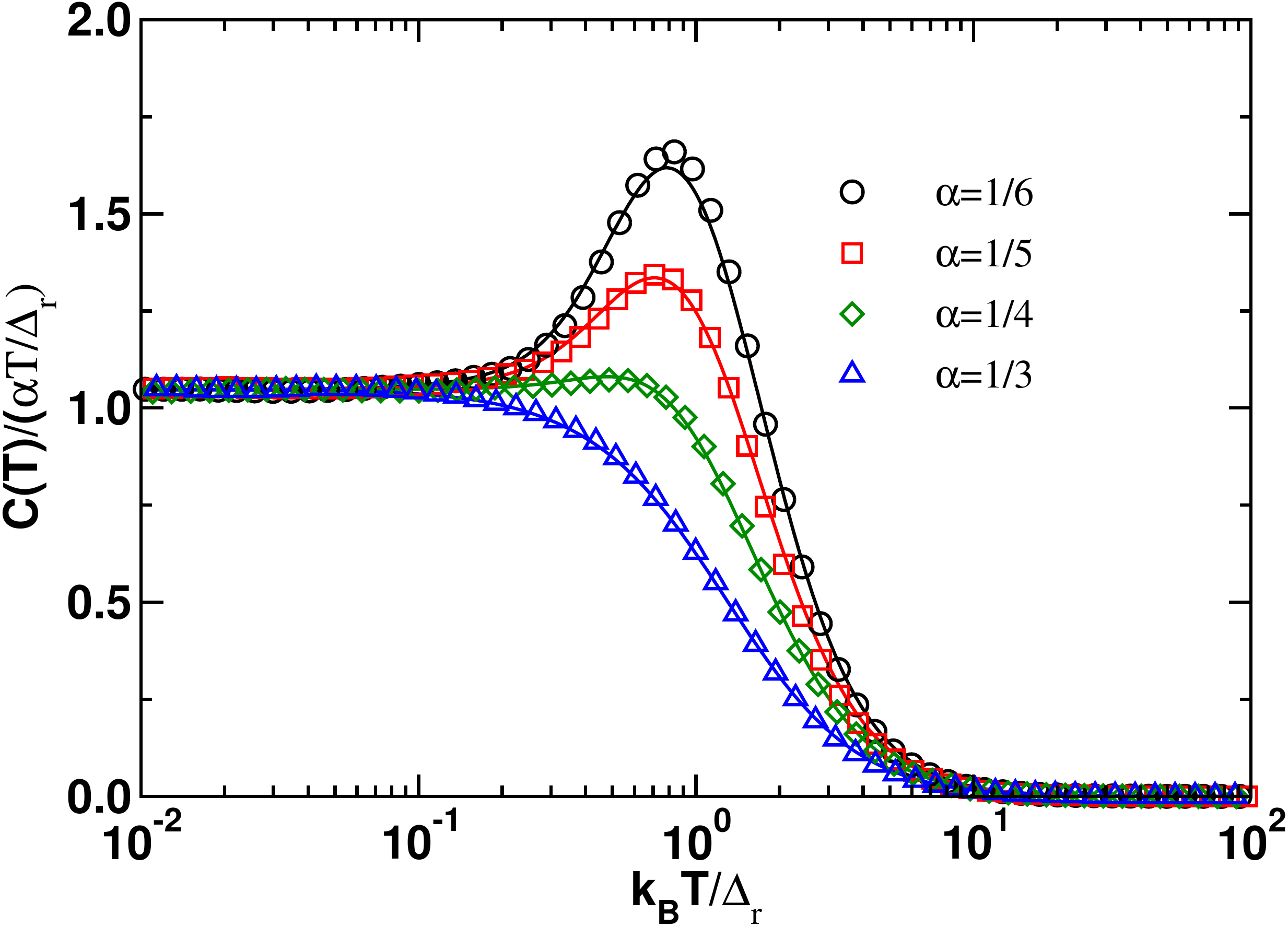}
\caption{{\em (Color online)}
  Specific heat, $C_{\rm imp}(T)/(\alpha k_{\rm B}T/\Delta_{\rm r})$, of the Ohmic 
  two state system as a function of reduced temperature $k_{B}T/\Delta_{\rm r}$, 
  for a range of dissipation strengths $\alpha < 1/2$. 
  Symbols: NRG results in new approach. Lines: Bethe ansatz results.
  In the low temperature Fermi liquid regime, $T\ll \Delta_{\rm r}$, we have 
  $C_{\rm imp}(T)/(\alpha k_{\rm B}T/\Delta_{\rm r})=\tilde{\gamma}+\tilde{\beta}
  (T/\Delta_{\rm r})^{2}$, with $\tilde{\gamma}=\pi/3$ and the $T^{3}$ coefficient in $C(T)$ 
  changes sign for $\alpha<1/3$ (see Ref.~\onlinecite{Costi1999}). 
  Model parameters: $\Delta_{\rm 0}/\omega_{\rm c}=0.005$. NRG parameters:
  $\Lambda = 10, n_{z}=4$ retaining $860$ states per NRG iteration.
\label{fig13}}
\end{figure}

The specific heat, $C_{\rm imp}=\partial E_{\rm imp}/\partial T$, of the Ohmic two-state system is defined via an impurity
internal energy $E_{\rm imp}=E_{\rm total}-E_{\rm 0}$, where 
$E_{\rm total}=\langle H\rangle =\langle H_{\rm S}\rangle + \langle H_{\rm B}\rangle + 
\langle H_{\rm I}\rangle$ and $E_{\rm 0}=\langle H_{\rm B}\rangle_{\rm 0}
=\sum_{i}\omega_{i}\langle a^{\dagger}_{i}a_{i}\rangle_{\rm 0} + E_{\rm zp}= \int_{0}^{\omega_{c}} d\omega\, 
\omega\, n(\omega)\,g(\omega) + E_{\rm zp}$ where $n(\omega)=1/(e^{\beta\omega}-1)$ is the 
Bose distribution function and the zero point energy $E_{\rm zp}$ can be dropped, as it cancels
in the difference $\langle H_{\rm B}\rangle - E_{\rm 0}=E_{\rm B}-E_{\rm 0}$ appearing in $E_{\rm imp}$. 
Evaluating $E_{\rm B}-E_{\rm 0}$ and $E_{\rm I}=\langle H_{\rm I}\rangle$ following
the approach in Sec.~\ref{sec:new approach}, we find
\begin{eqnarray}
E_{\rm B}-E_{\rm 0} &=& \frac{1}{\pi}\int d\omega n(\omega){\rm Im}
\left[\chi_{zz}(\omega+i\delta)
\frac{\partial}{\partial\omega}\left(\omega\Gamma(\omega+i\delta)\right)
\right]\nonumber\\
&=& E_{\rm B}^{(1)} + E_{\rm B}^{(2)},\label{band-contribution-sb}\\
E_{\rm B}^{(1)} &=&\frac{1}{\pi}\int d\omega n(\omega){\rm Im}
\left[\chi_{zz}(\omega+i\delta)\Gamma(\omega+i\delta)\right]\\
E_{\rm B}^{(2)} &=&\frac{1}{\pi}\int d\omega n(\omega){\rm Im}
\left[\chi_{zz}(\omega+i\delta)\omega\frac{\partial\Gamma(\omega+i\delta)}{\partial\omega}\right],
\end{eqnarray}
and
\begin{eqnarray}
E_{\rm I}&=& -\frac{1}{\pi}\int d\omega n(\omega){\rm Im}
\left[\chi_{zz}(\omega+i\delta)\Gamma(\omega+i\delta)\right],\label{int-contribution-sb}\\
\end{eqnarray}
where $\chi_{zz}(\omega+i\delta)=\langle\langle \sigma_{z};\sigma_{z}\rangle\rangle_{\omega+i\delta}$ 
is the longitudinal retarded dynamic susceptibility and 
$\Gamma(\omega+i\delta)$, characterizing the system-environment interaction, was defined above. 
Noting that 
$E_{\rm B}^{(1)}$ exactly cancels $E_{\rm I}$ in the impurity internal energy, we find
\begin{equation}
E_{\rm imp}= -\frac{1}{2}\Delta_{0}\langle\sigma_{x}\rangle 
+ \frac{1}{2}\epsilon\langle\sigma_{z}\rangle + E_{\rm B}^{(2)},
\end{equation}
that is, $E_{\rm imp}=E_{\rm S} + E_{\rm B}^{(2)}$. The term $E_{\rm B}^{(2)}$ gives
a non-negligible contribution to the impurity internal energy. For example, in the
Ohmic case with spectral function 
$J(\omega)=-\frac{1}{\pi}{\rm Im}\Gamma(\omega+i\delta)\sim \alpha\omega$ we have
$\omega\partial J(\omega)/\partial\omega ~\sim \alpha\omega$ at low frequencies, 
so $E_{\rm B}^{(2)}$ provides a contribution proportional to $\alpha$. 
By carrying out specific heat calculations on the AKM, we
find numerically that the impurity specific heat is consistent with setting
$E_{\rm B}^{(2)}=\frac{1}{2}\alpha\Delta_{0}\langle\sigma_{x}\rangle+A$, 
with $A$ being a weakly temperature dependent term, and negligible for calculating the
specific heat, except in the limit $\alpha\rightarrow 1^{-}$. The latter limit 
is difficult to treat numerically because of the vanishing low energy scale 
$\Delta_{\rm r}$ for $\alpha\rightarrow 1^{-}$ (e.g., for $\Delta_{\rm 0}/\omega_{\rm c}=0.01$ and
$\alpha=0.9$ we have $\Delta_{\rm r}/\omega_{\rm c}=10^{-20}$). 
Hence, except in this extreme limit, and as we show below by comparing with exact results,
the impurity specific heat can be obtained accurately from 
$C_{\rm imp}=\frac{\partial E_{\rm imp}}{\partial T}$ by using
\begin{equation}
E_{\rm imp}\approx -\frac{1}{2}\Delta_{0}(1-\alpha)\langle\sigma_{x}\rangle 
+ \frac{1}{2}\epsilon\langle\sigma_{z}\rangle\label{eq:energy-ohmic}.
\end{equation}
Figure~\ref{fig12} shows results obtained in this way for $C_{\rm imp}(T)/({k_{\rm B}T/\Delta_{\rm r}})$
compared to Bethe ansatz calculations for the AKM\cite{Costi1998} for a range of 
dissipation strengths. These results recover the known results for asymptotically high and
low temperatures.\cite{Goerlich1988} In common with specific heats of other correlated electron
systems as a function of interaction strength,\cite{Chandra1999} we observe a 
crossing point in $C(T)/T$ (here, at $k_{\rm B}T/\Delta_{\rm r} \approx 0.67$).
On decreasing the dissipation strength from strong ($\alpha>1/2$) to weak values ($\alpha<1/2$) the 
$T^{3}$ coefficient of the specific heat changes
sign for $\alpha<1/3$ resulting in the appearance of a finite temperature 
peak in $C(T)/T$. This is shown in more detail in Fig.~\ref{fig13}. It signifies the 
development of a gap $~\sim \Delta_{0}$ in the spectrum as $\alpha\rightarrow 0$. For $\alpha=0$
one eventually recovers the Schottky specific heat for a non-interacting two level system. The
expression (\ref{eq:energy-ohmic}) for the Ohmic two system is also the impurity internal energy 
of the equivalent AKM (indeed, the NRG results that we showed were for this model). The correspondence of 
model parameters was given above and the operators $\sigma_{x}$ and $\sigma_{z}$ are identified, 
under bosonization,\cite{Hakim1985,Leggett1987,Weiss2008,Costi1999} with the spin-flip operator 
$S^{+}s_{\rm 0}^{-}+S^{-}s_{\rm 0}^{+}$ and the local $S_{z}$ in the AKM, respectively. 
The zero temperature expectation values 
$\langle\sigma_{x}\rangle$ and $\langle\sigma_{z}\rangle$ (and the associated entanglement 
entropy of the qubit) have been studied previously as a function of dissipation strength and
finite bias.\cite{Costi2003,LeHur2008} 

We expect that the term $E_{\rm B}^{(2)}$ is non-negligible also for 
generic spectral functions $J(\omega)\sim \omega^{s}$ and certainly for
the sub-Ohmic case $s<1$.  Recent results for the local spin dynamics 
of the sub-Ohmic spin boson model \cite{Florens2011} could shed light on this. 

%\changes
{
The result (\ref{eq:energy-ohmic}) shows that a significant contribution 
to the impurity internal energy and specific heat arises from the (interacting) bath 
contribution $E_{\rm B}^{(2)}$, which remains finite for arbitrarily small 
$\alpha$. Thus, while a definition of the internal energy of the system  via 
$E_{\rm S}=\langle H_{\rm S}\rangle$ and the specific heat 
via $C_{\rm S}=\partial E_{\rm S}/\partial T$, might seem reasonable for a small quantum system 
{\em weakly} coupled to an infinite bath, such a definition yields,
in general, a specific heat $C_{\rm S}$ which differs from $C_{\rm imp}$.\cite{Haenggi2006,Haenggi2008,Ingold2009,Ingold2012}
One system for which the two definitions agree is the harmonic oscillator coupled Ohmically to an infinite
bath of harmonic oscillators.\cite{Haenggi2006} This result, however, represents a special case, and, moreover, 
is sensitive to details
of the cut-off scheme used for the spectral function $J(\omega)$ (see Ref.~\onlinecite{Haenggi2006,Ingold2012}).  
The use of $E_{\rm imp}$ and $C_{\rm imp}$ as definitions for the system internal energy 
and specific heat in the context of open quantum systems \cite{Weiss2008,Ford2005} also provides
an unambiguous prescription for their measurement in terms of two separate measurements,\cite{Ingold2009,Hasegawa2011} 
one for $H$ and one for $H_{0}$.

We note also that the impurity specific heat
$C_{\rm imp}(T)=C(T)-C_{0}(T)$ need not be positive at all temperatures and only the 
positivity of $C(T)$ and $C_{0}(T)$ in Eqs.~(\ref{spec-ctotal}) and (\ref{spec-chost})
is guaranteed by thermodynamic stability of 
the equilibrium systems described by $H$ and $H_{0}$ 
(see Ref.~\onlinecite{Callen1985}). Examples of systems where the
difference, $C_{\rm imp}(T)$, may be negative in some temperature range, 
include quantum impurities exhibiting a flow between a stable and an unstable fixed
point,\cite{Florens2004} and magnetic impurities in superconductors.\cite{Zitko2009} 
}

\section{Discussion and conclusions}
\label{sec:discussion and conclusions}
In this paper, we introduced a new approach to the calculation of impurity
internal energies and specific heats of quantum impurity models within 
the NRG method. For general Anderson impurity models, the impurity 
contribution to the internal energy was expressed in terms of local 
quantities and the main contribution to the impurity specific heat 
was shown to arise from local static correlation functions. For this class
of models,  the impurity specific heat can be obtained essentially exactly 
as $C_{\rm imp}(T) = \frac{\partial E_{\rm ionic}}{\partial T} 
+\frac{1}{2}\frac{\partial E_{\rm hyb}}{\partial T}$, where $E_{\rm ionic}=\langle
 H_{\rm imp}\rangle$ and $E_{\rm hyb}$ is the hybridization energy.
A comparison with exact Bethe ansatz calculations showed
that the results for specific heats of the Anderson impurity model are 
recovered accurately over the whole temperature and magnetic field range.
The new method has several advantages over the conventional approach to specific
heats within the NRG, namely, (i), only diagonalization of the total system
is required,  (ii), only local quantities are required, and, (iii),
discretization oscillations at large $\Lambda$ are significantly smaller 
than in the conventional approach.

For the dissipative two state system we obtain the specific heat as
$C_{\rm imp}(T)\equiv \frac{\partial E_{\rm imp}}{\partial T}=\frac{\partial E_{\rm S}}{\partial T}+\frac{\partial E_{\rm B}^{(2)}}{\partial T}$, 
where $E_{\rm S}=\langle  H_{\rm S}\rangle$ is analogous to 
$E_{\rm ionic}$ in the Anderson model, and $E_{\rm B}^{(2)}$ is a contribution 
to the energy of the system arising from the interaction with the bath. 
It depends on the local dynamical susceptibility and
the type of coupling to the environment. For the Ohmic case, we used the
equivalence of the Ohmic two state system to the AKM to show numerically that 
$E_{\rm B}^{(2)}=\frac{1}{2}\alpha\Delta_{0}\langle\sigma_{x}\rangle + A 
$ with $A$ having a negligible temperature dependence, 
except in the extreme limit $\alpha\rightarrow 1^{-}$.
Comparison with exact Bethe ansatz calculations on the AKM confirmed the above.

The approach described in this paper applies to energy dependent hybridizations
also, see Fig.~\ref{fig7}, so, inclusion of the term $E_{\rm int}^{(2)}$ in
Eq.~(\ref{eq:impurity-internal-energy-exact}), could prove useful in applications to quantum 
impurities with a pseudogap density of states.\cite{Gonzalez-Buxton1998,Vojta2002}
It may also be applied within other methods for solving quantum impurity models, 
for example, within continuous time \cite{Gull2011} or Hirsch-Fey \cite{Hirsch1986} quantum Monte Carlo 
techniques or exact diagonalization methods (for a recent review see Ref.~\onlinecite{Liebsch2012} 
and references therein). Local static correlation functions, such as the double occupancy,
required for $E_{\rm imp}$, are readily extracted within these approaches.\cite{Jakobi2009}

Within a DMFT treatment of correlated lattice models, 
\cite{Metzner1989,Georges1996,Kotliar2004,Vollhardt2012} the hybridization function $\Delta$
acquires an important temperature and frequency dependence $\Delta(\omega)\rightarrow 
\Delta(\omega,T)$. The latter enters explicitly in the term $E_{\rm int}^{(2)}$, whose
inclusion could offer an approach to the calulation of specific heats of correlated lattice
models. The thermodynamic potential of the latter \cite{Janis1991} is a sum of two parts,
one depending on the local self-energy, which is the central quantity calculated
in DMFT, and another equal to the thermodynamic potential, $\Omega_{\rm imp}=E_{\rm imp}-T S_{\rm imp}$, of
the effective impurity model. The latter can be obtained from 
$E_{\rm imp}(T)$, via $C_{\rm imp}(T)$ and $S_{\rm imp}(T)=\int_{0}^{T}dT' \frac{C_{\rm imp}(T')}{T'}$.
The impurity internal energy, expressed in terms of local dynamical quantities as in 
Ref.~\onlinecite{Kjoellerstroem1966}, has recently been used in a DMFT solution 
of the Hubbard model within a variational generalization \cite{Kauch2012} 
of the local moment approach.\cite{Logan1998}

In the future, it may be interesting, especially in the context of qubits or nanodevices, 
to consider the time dependence of the impurity internal energy subject to an initial 
state preparation, for example, within techniques such as time-dependent density matrix 
renormalization group \cite{Daley2004,White2004,Silva2008} or time-dependent NRG.\cite{Costi1997,Anders2005,Rosch2012}

\begin{acknowledgments}
We thank D. P. DiVincenzo, A. Rosch, S. Kirchner, A. Weichselbaum, 
G.-L. Ingold, P. H\"{a}nggi and A. Liebsch
for useful discussions and comments on this work, and A. Kauch for drawing our 
attention to Ref.~\onlinecite{Kjoellerstroem1966}. 
We acknowledge supercomputer support by the John von
Neumann institute for Computing (J\"ulich).
\end{acknowledgments}
\appendix
\section{Band contribution to impurity internal energy} 
\label{sec:appendix band contribution}
The expression (\ref{band-contribution}) for the conduction band 
contribution to the impurity internal energy 
requires evaluation of the integral
\begin{equation}
I(\omega) = \int d\epsilon \frac{\epsilon V^{2}N(\epsilon)}{(\omega-\epsilon+i\delta)^{2}}.
\end{equation}
We assume a density of states $N(\omega)$ vanishing at the band
edges at $\omega=\pm D$. The hybridization function $\Delta(\omega)=\sum_{k}V^{2}/(\omega-\epsilon_{k}+i\delta)=\Delta_{R}(\omega) + i\Delta_{I}(\omega)$ where $\Delta_{I}(\omega) =-\pi N(\omega) V^{2}$. With 
these definitions, we have
\begin{eqnarray}
I(\omega) &=& -\frac{1}{\pi}\int_{-D}^{+D}d\epsilon \epsilon\Delta_{I}(\epsilon)\frac{\partial}{\partial\epsilon}\frac{1}{(\omega-\epsilon+i\delta)}\nonumber\\
&=& -\frac{1}{\pi}\frac{\epsilon \Delta_I(\epsilon)}{\omega-\epsilon+i\delta}|_{-D}^{+D}
\nonumber\\& & +\frac{1}{\pi}\int_{-D}^{+D} d\epsilon \frac{1}{\omega-\epsilon+i\delta}\frac{\partial}{\partial\epsilon}(\epsilon \Delta_{I}(\epsilon)).
\end{eqnarray} 
The first term vanishes since $\Delta_{I}(\pm D)=0$ 
for regular (e.g., $3D$) densities of states (and will otherwise result in 
contributions with negligible temperature dependence). The second term can be evaluated by
noting that $\Delta(\omega+i\delta)$ satisfies the causal properties of 
retarded Green functions and by using the following properties of principle value (P.V.) integrals: 
if P.V.[$f(x)$] $=g(y)$ then P.V.[$f'(x)$] $=g'(y)$ and P.V.[$xf(x)$] $=yg(y)+\frac{1}{\pi}\int dx f(x)$.
The final result is 
\begin{equation}
I(\omega) = -\frac{\partial}{\partial\omega}(\omega \Delta(\omega))
\end{equation} 
\section{Numerical solution of the Thermodynamic Bethe Ansatz equations}
\label{sec:appendix tba calculations}
\def\convolution{*}
In this Appendix, we summarize the thermodynamic Bethe ansatz (TBA) equations for the Anderson model,
which were derived by Okiji and Kawakami \cite{Kawakami1981,Kawakami1982a,Okiji1983} and Tsvelick, Filyov,
and Wiegmann, \cite{Wiegmann1983,Tsvelick1983,Filyov1982,Tsvelick1982} and provide details of their
numerical solution.\cite{Rajan1982,Desgranges1985,Sacramento1991,Costi1999,Takahashi2002,Bolech2005}
The numerical procedure described applies to both the symmetric and asymmetric Anderson models and in the 
presence of a finite magnetic field and was used to obtain the results presented in this paper.

\subsection{Thermodynamic Bethe Ansatz Equations}  
The thermodynamic Bethe ansatz (TBA) produces an infinite set of coupled integral equations for the functions 
$\epsilon(k)$, $\kappa_{n}'(\Lambda)$ and $\kappa_{n}(\Lambda)$, $n=1,2,\dots$, 
describing the charge and spin excitations of the system 
(Tsvelick and Wiegmann~\cite{Tsvelick1982}):
\begin{subequations}
\begin{align}
 \epsilon(k) & -  T \cdot \int_{-\infty}^{\infty} s(g(k)-\Lambda) \cdot \ln(f(\kappa_1(\Lambda))\mathrm{d}\Lambda 	= \nonumber \\
& \epsilon_0(k) -  T \cdot\int_{-\infty}^{\infty} s(g(k)-\Lambda) \cdot \ln(f(\kappa'_1(\Lambda))\mathrm{d}\Lambda \\
 \kappa_n(\Lambda) & + T \cdot (s\convolution (\ln(f(\kappa_{n+1})) + \ln(f(\kappa_{n-1}))))(\Lambda) = \nonumber \\
& \delta_{n,1} \cdot T \cdot \int^\infty_{-\infty} s(g(k)-\Lambda) \cdot \ln(f(-\epsilon(k)))\cdot g'(k)\, \mathrm{d}k  \\
 \kappa'_n(\Lambda) & + T \cdot (s\convolution (\ln(f(\kappa'_{n+1})) + \ln(f(\kappa'_{n-1}))))(\Lambda) = \nonumber \\
& \delta_{n,1} \cdot T \cdot \int^\infty_{-\infty} s(g(k)-\Lambda) \cdot \ln(f(\epsilon(k)))\cdot g'(k)\, \mathrm{d}k 
\label{bethe-ansatz:1stchange}
\end{align}
\end{subequations}
where 
\begin{align*}
 & g(k) = \frac{(k - \varepsilon_{d} - \frac{1}{2}U)^2}{2 \Gamma \mathrm{U}}, \quad
   s(\Lambda) = \frac{1}{2 \cosh(\pi \Lambda)},\\
&   f(k) = \frac{1}{1 + \mathrm{e}^{k/T}}, \quad
  R(x) = \frac{1}{\pi}\int_{0}^\infty \frac{\cos(\omega x) }{1 + \mathrm{e}^\omega} \mathrm{d}\omega \\
&  \epsilon_0(k) = k - \varepsilon_{d} - \frac{1}{2}U + \int_{-\infty}^\infty R(g(k) -g(p))\cdot p\cdot g'(p) \mathrm{d}p
\end{align*}
$g'(k)$ denotes the first derivative of $g(k)$ with respect to $k$.  $\convolution$ is the convolution of two functions. $\kappa_0$ and $\kappa'_0$ equal $-\infty$. For $n\rightarrow\infty$ the functions approach the constant values,
\begin{align}
 \lim\limits_{n\rightarrow\infty} \kappa_n = n\cdot H,\qquad \lim\limits_{n\rightarrow\infty} \kappa'_n = n\cdot (2 \varepsilon_{d} + U),
\label{bethe-ansatz:2ndchange}
\end{align}
where $H$ is a uniform magnetic field and  $2 \varepsilon_{d} + U$ measures the deviation from the
symmetric point at $\varepsilon_{d}=-U/2$.
The impurity contribution to the specific heat, $C_{\rm imp}$, may be calculated from the
the impurity contribution to the thermodynamic potential, $\Omega_{\rm imp}$, via
$C_{\rm imp}=-T\partial^{2}\Omega_{\rm imp}/\partial T^{2}$, where
\begin{align}
 \Omega_{\rm imp}  =  & T \int_{-\infty}^\infty \rho_0(k) \cdot \ln(f(-\epsilon(k))) \mathrm{d} k \nonumber \\
	    & + T \int_{-\infty}^\infty \sigma_0(\Lambda) \cdot \ln(\kappa'_1(\Lambda)) \mathrm{d} \Lambda + E_{\rm 0}
\end{align}
The functions $\rho_0$ and $\sigma_0$ are given by:
\begin{align*}
 \sigma_0(\Lambda) & = \int_{-\infty}^{\infty} s(\Lambda - g(k))\cdot\Delta(k) \mathrm{d}k\\
 \rho_0(k) & = \Delta(k) + g'(k) \cdot \int_{-\infty}^{\infty} R(g(k) -g(p))\cdot\Delta(p)\mathrm{d}p,
\end{align*}
where $ \Delta(k) =  \frac{\Gamma}{\pi (\Gamma^2 + (k - \varepsilon_{d})^2)}$.
$E_{\rm 0}$ is the ground state energy of the symmetric Anderson model.\cite{Kawakami1981,Kawakami1981}
Note two changes with respect to the earlier Ref.~\onlinecite{Tsvelick1982}: 
a sign change in equation~\ref{bethe-ansatz:1stchange} (as in Wiegmann 
and Tsvelick~\cite{Wiegmann1983})  and a factor $2$ in the boundary value 
for $\kappa'_n$ in equation~\ref{bethe-ansatz:2ndchange} 
(as in Okiji and Kawakami~\cite{Okiji1983}). 

For the calculations we use a transformation of $\kappa_{n}$ and $\kappa_{n}'$ 
to new functions $\xi_{n}$ and $\xi_{n}'$ similar 
to that used in previous works.\cite{Rajan1982,Desgranges1985,Costi1999}
After substituting $\xi_n = \ln(1 + \mathrm{e}^{\kappa_n/T})$ and $\xi'_n = \ln(1 + \mathrm{e}^{\kappa'_n/T})$ we obtain 
the following coupled equations:
\begin{subequations}
\begin{align}
 \xi_1(\Lambda) &= \ln(1 + \exp((s\convolution(\xi_{2} + I_1))(\Lambda)))\label{eq:ksi1} &\\
 \xi_n(\Lambda) &= \ln(1 + \exp((s\convolution(\xi_{n-1} + \xi_{n+1}))(\Lambda)))\\
 \xi'_1(\Lambda) &= \ln(1 + \exp((s\convolution(\xi'_{2} + I'_1))(\Lambda)))\label{eq:ksi1'} &\\
 \xi'_n(\Lambda) &= \ln(1 + \exp((s\convolution(\xi'_{n-1} + \xi'_{n+1}))(\Lambda)))\\
 I_1(\Lambda) &= \int^\infty_{-\infty} s(g(k)-\Lambda) \cdot \ln(f(-\epsilon(k)))\cdot g'(k)\, \mathrm{d}k& \\
 I'_1(\Lambda) &= \int^\infty_{-\infty} s(g(k)-\Lambda) \cdot \ln(f(\epsilon(k)))\cdot g'(k)\, \mathrm{d}k& \\
 I(k) & = \int_{-\infty}^{\infty} s(g(k)-\Lambda) \cdot (\xi_1(\Lambda) - \xi'_1(\Lambda))\mathrm{d}\Lambda& \\
 e(k) &= e_0(k) +  T \cdot I(k)&
\end{align}
\end{subequations}

\subsection{Truncation}
\begin{figure}
\centering
 \includegraphics[width=\linewidth,clip]{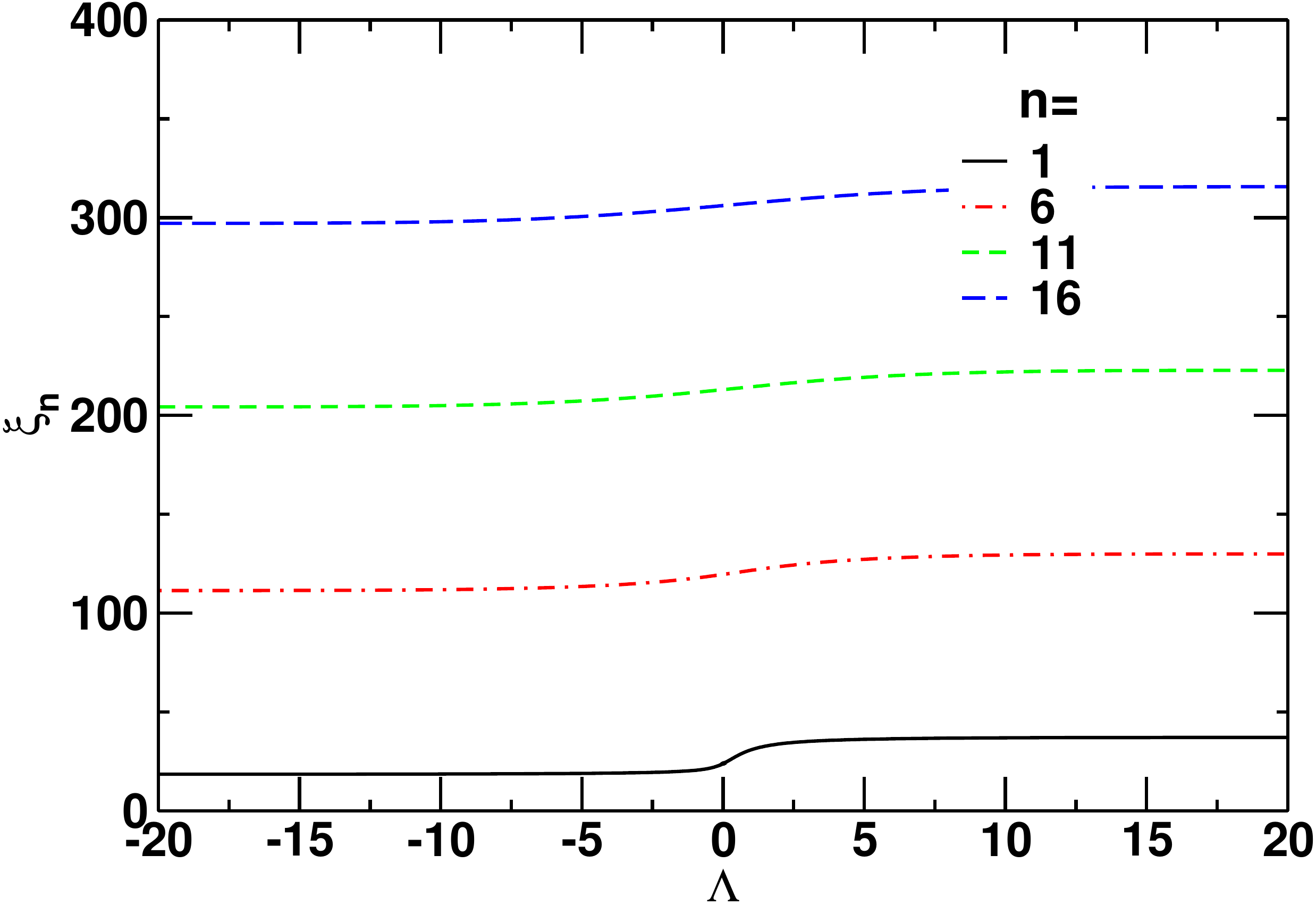} 
 \caption{
{\em (Color online)}
The figure shows a set of $\xi_n$ for the symmetric case ($\varepsilon_{d} + U/2 =0$) 
zoomed to range of $\Lambda=-20 \dots 20$. The functions become smoother 
with higher $n$ due to the convolution with $s(x)$.
}
\label{fig14}
\end{figure}
For calculational purposes the equations $\xi_n$ 
and $\xi'_{n'}$ have to be truncated at some finite value $n=N$ and $n'=N'$. 
One has to calculate the functions at the truncation with care, to avoid 
wrong results at the boundaries $\Lambda\rightarrow \pm \infty$. 
We use the truncation scheme of Takahashi and Shiroishi.\cite{Takahashi2002}
It is assumed that the function $s(x)$ can be approximated by 
$\delta(x)/2$ for large $n$ or $n'$. This is justified as the functions 
become smoother in this region (see figure~\ref{fig14}). 
Rewritten for the Anderson Model and for $\xi_N$ and $\xi'_N$ the corresponding 
truncation functions are calculated by:
\begin{subequations}
\label{bethe-ansatz:bounds}
\begin{align}
  \xi_{N} &=  \ln \Big(\big(\cosh(\frac{H}{2})\cdot\sqrt{2 + \mathrm{e}^{\xi_{N-1}}} + \nonumber \\
  & \qquad \sqrt{1 + \sinh^2(\frac{H}{2}) \cdot [2 + \mathrm{e}^{\xi_{N-1}}]}\big)^2\Big) \\
 \xi'_{N'} &= \ln \Big(\big(\cosh(\frac{2 \varepsilon_{d} + U}{2})\cdot\sqrt{2 + \mathrm{e}^{\xi'_{N'-1}}} + \nonumber \\ 
  &\qquad \sqrt{1 + \sinh^2(\frac{2 \varepsilon_{d} + U}{2}) \cdot [2 + \mathrm{e}^{\xi'_{N'-1}}]}\big)^2\Big)
\end{align}
\end{subequations}

As a further check, and to ensure the correct behaviour at the boundaries, the TBA integral equations 
were explicitly solved in the limits of 
$\Lambda, k \rightarrow \pm \infty$.
As the functions are smooth in this limit one can assume that
$s(x) \rightarrow \delta(x)/2$ and $\lim\limits_{k\rightarrow \infty}\epsilon_0(k) 
= 2(k - \varepsilon_{d} - U/2)$, $\lim\limits_{k\rightarrow -\infty}\epsilon_0(k) = 0$. 
This leads to the following set of coupled algebraic equations:
\begin{subequations}
\begin{align}
 \lim\limits_{\Lambda\rightarrow -\infty} &&\nonumber \\
 \xi_1  &= \ln(1 + \exp(\frac{1}{2}\xi_2))) &\\
 \xi_n  &= \ln(1 + \exp(\frac{1}{2}(\xi_{n-1}+\xi_{n+1}))) &\\
 \xi'_1 &= \ln(1+\exp(\frac{1}{2}\xi'_2))& \\
 \xi'_n  &= \ln(1 + \exp(\frac{1}{2}(\xi'_{n-1}+\xi'_{n+1}))) &\\
 \lim\limits_{\Lambda\rightarrow \infty} &&\nonumber \\
 \xi_1 & = \ln(1 + \exp(\frac{1}{2}(\xi_2 - \ln(1 + \exp(\frac{1}{2}\xi_1)))))& \\
 \xi_n  &= \ln(1 + \exp(\frac{1}{2}(\xi_{n-1}+\xi_{n+1}))) &\\
 \xi'_1 & = 0 & \\
 \xi'_n  &= \ln(1 + \exp(\frac{1}{2}(\xi'_{n-1}+\xi'_{n+1})))&
\end{align}
\end{subequations}
The truncation constants $\xi_N$ and $\xi'_{N'}$ are calculated as 
in equation~\ref{bethe-ansatz:bounds}. The boundary values where calculated 
by iteration using a modification of the Powell hybrid method.

\subsection{Numerical Details}
\begin{figure}
\includegraphics[width=\linewidth,clip]{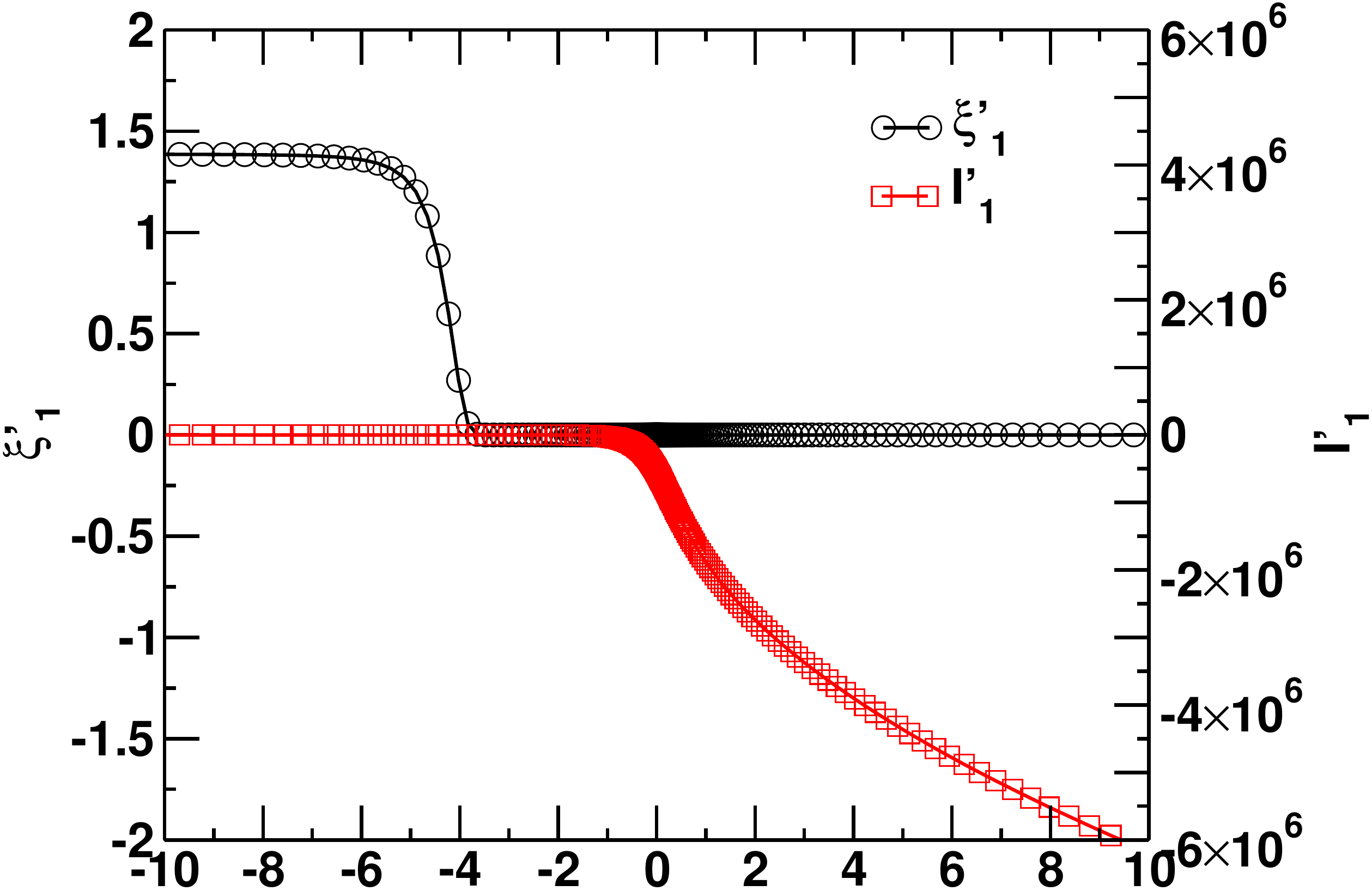}
\caption
{{\em (Color online)}
  Comparison between $I'_1$ and $\xi'_1$ for 500 iterations and $T= 10^{-4} T_K$. 
  Parameters were chosen to be the same as in figure~\ref{fig1}.
  For very low temperatures $\xi'_1$ (circles, left $y$-axis) 
exhibits an exponential drop beyond a certain
rapidity $\Lambda_{0}$ ($\approx -4$ for the case shown) 
which is difficult to capture with a fixed grid. This problem can be overcome by using 
the smooth function $I_{1}'$ (squares, right $y$-axis) to calculate $\xi_{1}'$ 
via Eq.~(\ref{eq:ksi1'}).
}
\label{fig15}
\end{figure}

For the calculations, a logarithmic grid was used that is centred 
around $\varepsilon_{d} + U/2$. The TBA equations were solved by iteration. The 
initial values of $\xi_n$ and $\xi'_{n'}$  were chosen to fit a $\tanh$-function 
with boundary values given by the correct boundary values 
of $\xi_n$ and $\xi'_n$, obtained as described above. The integrations were carried out 
using adaptive routines with the integrands being represented by 
splines of smooth functions only (see below). A smoother convergence of the iteration
procedure is obtained by using $10\%$ of the old iteration 
values in each step. To represent only smooth functions as splines, 
$\xi_1$ and $\xi'_1$ are not interpolated, but instead the $s\convolution \xi_2$ and 
$s\convolution \xi'_2$ respectively. The values of $\xi_1$ and $\xi'_1$ are 
then calculated from these convolutions and from $I_1$ and $I'_1$ using 
Eq.~(\ref{eq:ksi1}) and Eq.~(\ref{eq:ksi1'}). This avoids numerical problems due
to the exponential drop to zero of $\xi_{1}'$ beyond a certain rapidity $\Lambda_{0}$. 
See Figure~\ref{fig15} for a comparison between the behavior of 
$\xi'_1$ and $I'_1$.
$N=N'=20$ functions were used and iterated $500$ 
times for the figures in this section (and $2000$ times
for the results in the paper). 
The growth-rate of the grid was $1.05$ and it consisted of 801 points. The mid 400 values 
lie in a range of $[-40,40]$. After a certain temperature dependent cut-off 
($\pm 40 \pm 40 \cdot T/T_0$) the boundary values were used instead of being calculated 
to ensure numerical stability. The thermodynamic potential was calculated in a range of 
$T_0 \cdot 10^{-3}$ to $T_0 \cdot 10^6$ on a logarithmic mesh (factor $2^{1/8}$ as step 
width) where $T_0$ is defined as 
$T_0=\sqrt{U \Gamma/2}\cdot \exp(-\frac{\pi U}{8\Gamma} + \frac{\pi \Gamma}{2 U})$,  
Kondo temperature for the symmetric case. It is related to the magnetic 
susceptibility at zero temperature $\chi_{\rm imp}(T=0)=\frac{(g\mu_{\rm B})^2}{4 k_{\rm B} T_0}$
(see Hewson in Ref.~\onlinecite{Hewson1997} p. 165, and Kawakami and Okiji 
in Ref.~\onlinecite{Kawakami1982b}). 

\bibliography{spec}
\end{document}